\documentclass{raa}

\usepackage{graphicx,times,natbib}             %for PS/EPS graphics inclusion, new
\input{epsf.sty}                        %for PS/EPS graphics inclusion, old
\usepackage{subfig}

\begin{document}

   \title{The Tidal Tails of Globular Cluster Palomar 5 Based on
   Neural Networks Method
%\footnotetext{$*$ Supported by the National Natural Science Foundation of China.}
}
%   \subtitle{I. Place Your Subtitle Here}

   \volnopage{Vol.0 (200x) No.0, 000--000}      %%preserved for Editor. DOn't remove!
   \setcounter{page}{1}          %%starting page, preserved for Editor. DOn't remove!

   \author{H. Zou
        \inst{1,2}
        \and Z. -Y. Wu
        \inst{1}
        \and J. Ma
        \inst{1}
        \and X. Zhou
        \inst{1}
        }

   \institute{National Astronomical Observatories, Chinese Academy of Sciences,
             Beijing 100012, China; {\it zywu@bao.ac.cn} \\
             \and
             Graduate University of Chinese Academy of Sciences, Beijing 100049, China}

   \date{Received~~2009 month day; accepted~~2009~~month day}

\abstract{The Sixth Data Release (DR6) in the Sloan Digital Sky
Survey (SDSS) provides more photometric regions, new features and
more accurate data around globular cluster Palomar 5. A new method, Back Propagation
Neural Network (BPNN), is used to estimate the probability of
cluster member to detect its tidal tails. Cluster and field stars,
used for training the networks, are extracted over a $40\times20$
deg$^2$ field by color-magnitude diagrams (CMDs). The best BPNNs
with two hidden layers and Levenberg-Marquardt (LM) training
algorithm are determined by the chosen cluster and field samples.
The membership probabilities of stars in the
whole field are obtained with the BPNNs, and contour maps of the
probability distribution show that a tail extends $5.42\dg$ to the north
of the cluster and a tail extends $3.77\dg$ to the south. The whole
tails are similar to those detected by \citet{od03}, but no
longer debris of the cluster is found to the northeast of the sky.
The radial density profiles are investigated both along the tails
and near the cluster center. Quite a few substructures are
discovered in the tails. The number density profile of the cluster is fitted
with the King model and the tidal radius is determined as $14.28'$.
However, the King model cannot fit the observed profile at the outer regions ($R > 8'$) because of the
tidal tails generated by the tidal force. Luminosity functions of the cluster and the tidal tails are
calculated, which confirm that the tails originate from Palomar 5.
}

   \authorrunning{H. Zou, Z.-Y. Wu, J. Ma, X. Zhou}            %author_head in even pages
   \titlerunning{Tidal tail of Palomar 5}  % title_head in odd pages

   \maketitle
%% The author head (on even pages) and the title head (on odd pages) will be
%% automatically extracted from \author{} and \title{}. Whenever the title is too long,
%% you will be asked to supply a shorter one by inserting either \authorrunning{} or
%% \titlerunning{} before \maketitle. Anyway, you can specify your own heads.
%%
%%
%% Note: In the following text body of your manuscript, please note several differences from
%%       other major journals:
%% (1) \subsection{Please Capitalize the First Letter of Each Notional Word in Subsection Title}
%% (2) Please Capitalize the First Letter of Each Notional Word in all tables' captions

%
%________________________________________________ sections below
%
\section{Introduction}           %% first-level sections will be auto-capitalized
\label{intro} Globular clusters (GCs) are the oldest populations in
the Galaxy. Most of  GCs, which formed in the early days of the
Galaxy, have been destroyed by various mechanisms \citep{wu03}. Mass
loss from stellar evolution is very important during the first $\sim
1$ Gyr of cluster evolution, and most of low mass clusters have been
dissolved during this early phase. For survival clusters, their
evolutions will be dominated by the internal dynamical processes
caused by encounters between cluster stars (the two-body relaxation)
\citep{sp87}. GCs in the Galaxy have elliptical orbits, and some of
them can move into the central region of the Galaxy with
perigalactic distances less than $\sim 1$ kpc \citep{wu04}. When a
cluster crosses the bulge or disk of the Galaxy with timescale
shorter than its internal dynamical time, the cluster stars will
gain energy and speed up the evaporation. Such an interaction is
referred to as the tidal shock \citep{sp87}. Stars evaporating from
the cluster due to two-body relaxation or tidal shocks, will not
leave the cluster and merge into the Galactic field immediately.
They will move along the same orbit of the cluster and form the
`tidal tail' of the cluster.

\cite{gr95} examined the outer structures of 12 Galactic globular
clusters using star-count analysis with deep, two-color photographic
photometry. They found that most of their sample clusters show
extra-tidal wings in their surface density profiles. Two-dimensional
surface density maps for several clusters indicate the expected
appearances of tidal tails. \cite{le00} used large-field
photographic photometry of 20 globular clusters to investigate the
presence of tidal tails around these clusters; in this study,
star-count analysis and wavelet transform were used to detect the
weak structures formed by the stars that previously are members of
the clusters; and most of globular clusters in their sample display
large and extended tidal tails, which exhibit projected directions
towards the Galactic center.

The studies of \cite{gr95} and \cite{le00} are all based on
photographic observations covering large areas around the clusters.
The low signal-to-noise in the photographic photometry and serious
contaminants from the background galaxies make the detected tidal
tails in some clusters uncertain. The SDSS can provide large, deep
CCD imaging in five passbands coving 10,000 deg$^2$ in the sky. The
SDSS can also separate the stars and galaxies very well and is very
efficient in detecting tidal tails around globular clusters in the
Galaxy. Using SDSS data, well-defined tidal tails in some globular
clusters have been identified: Palomar 5 \citep{od01,od03,gr06b},
NGC5466 \citep{be06,gr06a}, and NGC 5053 \citep{la06}.

In above mentioned studies, Palomar 5 is a prominent object. It is a
remote globular cluster located at a distance of 23.2 kpc from the
Sun and has a tidal radius of  about  16.3 arcmin \citep{ha96}.
Using the ESO and SERC survey plates covering about $2.0\dg \times
2.0\dg$  in $R$ and $J$ filters, \cite{le00} searched the tidal
tails around this cluster and found that the detected structures
outside this cluster are strongly biased by the background galaxy
clusters appearing in the field, and it is difficult to derive any
conclusions on the genuine location of stars in the tidal tails of
this cluster.

Using SDSS data concentrating on Palomar 5 in a region with right
ascensions $226\dg \leq \alpha \leq 232\dg$ and declinations
$-1.25\dg \leq \delta \leq +1.25\dg$, \cite{od01} searched the tidal
tails of this cluster based on the empirical photometric filtering
method of \cite{gr95}. They found two well-defined tidal tails
emerging from this cluster to stretch out symmetrically to both
sides of the cluster and extend an angle of $2.6\dg$ on the sky.
Using the new SDSS data (before the public data release DR1)
yielding complete coverage of a region with a $6.5\dg$ to $8\dg$
wide zone along the equator and right ascension from $224\dg$ to
$236\dg$, and based on optimal contrast filtering method,
\cite{od03} found that the tidal tails of Palomar 5 have a much
larger spatial extent and can be traced about an arc of $10\dg$ on
the sky. More recently, using the SDSS DR4 data, \cite{gr06b}
applied the optimal contrast filtering method of \cite{od03} to
tracing the tidal tails of Palomar 5 in a region $224\dg <\alpha <
247\dg$ and $-3\dg < \delta < +10\dg$, and found the tidal tails to
extend to some $22\dg$ on the sky.

Most of the studies are based on star-count analysis in
color-magnitude space. While \citet{be06} gave us a brand new angle
of view to recognize the tidal tails of clusters. Their method is
based on an intelligent computing technique, Artificial Neural
Networks (ANN), which has been applied in many sorts of areas, such
as classification and pattern recognition. In the study of
\citet{be06}, back propagation neural network --- the most widely
applied ANN --- was used to estimate the probability of cluster
member for each object in the SDSS 5-band data space. Compared to
\citet{od03}, BPNN makes full use of the photometric information,
not just one CMD. Therefore, in this paper, we will introduce this
method to investigate the tidal tails of Palomar 5, where the DR6
data in a larger region ($40\times20$ deg$^2$) are used.

In $\S\ref{data}$, we describe the details of the SDSS DR6 and
preprocessing of the observed data. Section \ref{BPNN} presents the
general idea of BPNN. In $\S\ref{appl}$, we construct BPNNs with the best performances
after being trained with properly selected training data, and then
we apply them to the tidal tails detection of Palomar 5. Section
\ref{diss} discusses the profiles and features of the tails. A brief
conclusion is given in $\S$\ref{conc}.

\section{The Star Sample}\label{data}
\subsection{Observations}
In this study, the photometric data in the SDSS DR6 are used. The SDSS is a
photometric and spectroscopic survey, providing detailed optical
images covering more than a quarter of the sky and a 3-dimensional
map of about a million galaxies and quasars. A dedicated, 2.5-meter
telescope is located on Apache Point, New Mexico, equipped with a
120-megapixel camera and a pair of spectrographs fed by optical
fibers measuring more than 600 sources in a single observation.
There are 30 photometric CCDs with size $2048\times2048$ pixels for
each. The field of view is $3.0\dg$, and 5 broad band filters with
the wavelength ranging from 3000 \AA~to 10000 \AA~are used when
photometric images are taken. By far, subsequent data releases have
been published, including Early Data Release (EDR), DR1, DR2, DR3,
DR4, DR5, DR6 and DR7.

DR6 \citep{ad08} is the first release which has significant
changes about the processing software since DR2. For example,
calibrations are improved using cross-scans to tie the photometry of
the entire survey to each other. The photometric calibration is
improved with uncertainties of roughly $1\%$ in $g$, $r$, $i$ and $z$, and
$2\%$ in $u$, which are substantially better than the ones in previous
data releases. In addition, the magnitude limits are 22.0 for $u$,
22.2 for $g$ and $r$, 21.3 for $i$, and 20.5 for $z$. More
importantly, compared with DR4, DR6 includes new observed regions
where we can search the tidal tails of Palomar 5.

\citet{od03} only considered a limited region covering an area of 87
deg$^2$ and found the tails extending about an angular distance of
$10\dg$. No further investigation in the north, where photometric
data exist, was made to see whether the tails are longer.
\citet{gr06b} detected a $22\dg$ tidal tails in a larger region. Far
from the center of Palomar 5, these newly discovered tails do not
appear clearly, since signals and the background noise are so
similar. On the other hand, \citet{gr06b} used the DR4, in which a
narrow strip ($\alpha > ~228.5\dg$ and $0.5\dg < \delta < 1.5\dg$)
stretching to the north tidal tail has no photometric data, while
DR6 supplements this area. Therefore, in an area of $40\times20$
deg$^2$ ($220\dg < \alpha < 260\dg$ and $-5\dg < \delta < 15\dg$),
DR6 provides more photometric data. Due to tremendous number of
objects included in the area, we partition the whole field into 4
regions equally: R1 (R.A.: $220\dg \sim 230\dg$, Dec.: $-5\dg \sim
15\dg$); R2 (R.A.: $229\dg \sim 240\dg$, Dec.: $-5\dg \sim 15\dg$);
R3 (R.A.: $239\dg \sim 250\dg$, Dec.: $-5\dg \sim 15\dg$) and R4
(R.A.: $249\dg \sim 260\dg$, Dec.: $-5\dg \sim 15\dg$), where any
two contiguous regions are overlapped by an area of $1\times20$
deg$^2$ in order to avoid bad smoothing at the edges (see
$\S\ref{appl}$).

The information, which we need to detect the tidal tails of Palomar
5, includes: coordinate (J2000), point spread function (PSF)
magnitude ($r_{\textrm{psf}}$) and exponential model magnitude
($r_{\textrm{exp}}$) of $r$ band, the reddening values and the type
of each source. The way to separate stars from galaxies, the
definition of $r_{\textrm{psf}}$ and $r_{\textrm{exp}}$, photometric
and astrometric data reduction, and relative information can be
referred to the works \citep{lu01,st02,ab04}. We use so-called
Cmodel magnitudes (the default provided values of $ugriz$) as our
photometric data, because Cmodel magnitude is the best fit of
exponential and de Vaucouleurs models in each band, and it agrees
excellently with both PSF magnitude of stars and Petrosian magnitude
of galaxies. Even though we only extract stars to check the tidal
tails, it's impossible for us to promise that there are not any
miscellaneous galaxies which can not be distinguished by SDSS data
reduction piplines. Thus, for uniformity and the validity of the
photometry of both stars and galaxies, we choose the universal
magnitudes (Cmodel). Reddening corrections for each object are
deducted based on the reddening values from \citet{sc98}.

In our selected sky field, we obtain 15,305,060 sources in the
catalog, where there are 7,410,896 stars and 7,894,164 galaxies
classified by SDSS piplines. There are 5,458,077 sources in R1,
5,508,433 in R2, 4,164,575 in R3 and 173,975 in R4, respectively.

\subsection{Data Preprocessing}
\begin{figure}[h]
  \centering
    \epsscale{1}{1}
    \includegraphics[width=6in]{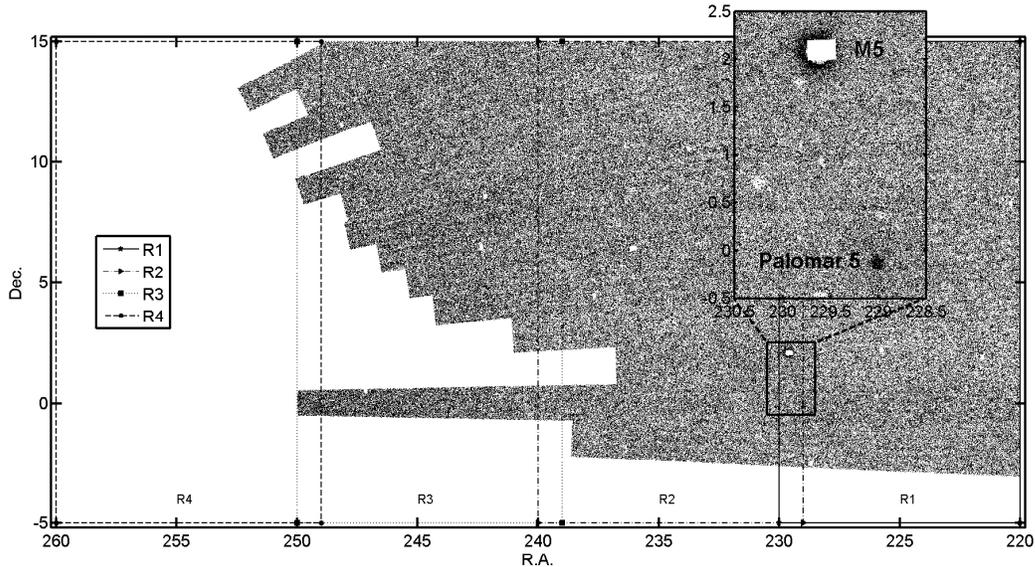}
    \caption{Distribution of all point sources with magnitude range $14 \leq r \leq
    22$ in the $40\times20$ deg$^2$ area. This area are divided into four
    overlapped regions (R1, R2, R3 and R4) to be processed in batches
    because of too large number of objects. M5 (center: $\alpha = 229.6\dg,
    \delta = 2.1\dg$) and Palomar 5 (center: $\alpha = 229.02\dg$, $\delta = -0.11\dg$)
    are two star-focus regions. The smaller box ($228.5\dg < \alpha < 230.5\dg,
    -0.5\dg < \delta < 2.5\dg$) encloses these two objects and the bigger one (subplot)
    shows the magnified image of the smaller one. The blank areas are regions which
    may be very dense clusters or bright stars or the regions the SDSS hasn't covered. }
    \label{f1}
\end{figure}

For source type determination in the SDSS, there are some flaws. For
example, occasionally SDSS piplines fail to distinguish blenders and
pairs of stars with small separations. Sometimes, the classification
scheme regards Seyfert galaxies or QSOs as stars \citep{st02}, and
overflows of very bright stars are identified as galaxies.
Furthermore, due to variations of observing conditions and natural
differences in diverse fields, the completeness of object detection
is fluctuant. Considering magnitude limits and the situations
depicted above, it is necessary to give cutoffs of magnitude to
avoid unnecessary impurities. So we select the sample stars with a
magnitude scope $14 \leq r \leq 22$. Fig. \ref{f1} shows a visual
impression of the selected star sample, and the photometric
boundaries. In Fig. 1, only $1/30$ sample stars are randomly
selected to be drawn in order not to be too black. M5 and Palomar 5
are also indicated in Fig. 1. The subplot in this figure is the part
including M5 and Palomar 5, which is enlarged. The galactocentric
distance of M5 is about 6.2 kpc, far away from Palomar 5, whose
galactocentric distance is about 18.6 kpc. So M5 hardly has effect
on Palomar 5.

\begin{figure}[h]
  \centering
    \epsscale{1}{1}
    \includegraphics[width=6in]{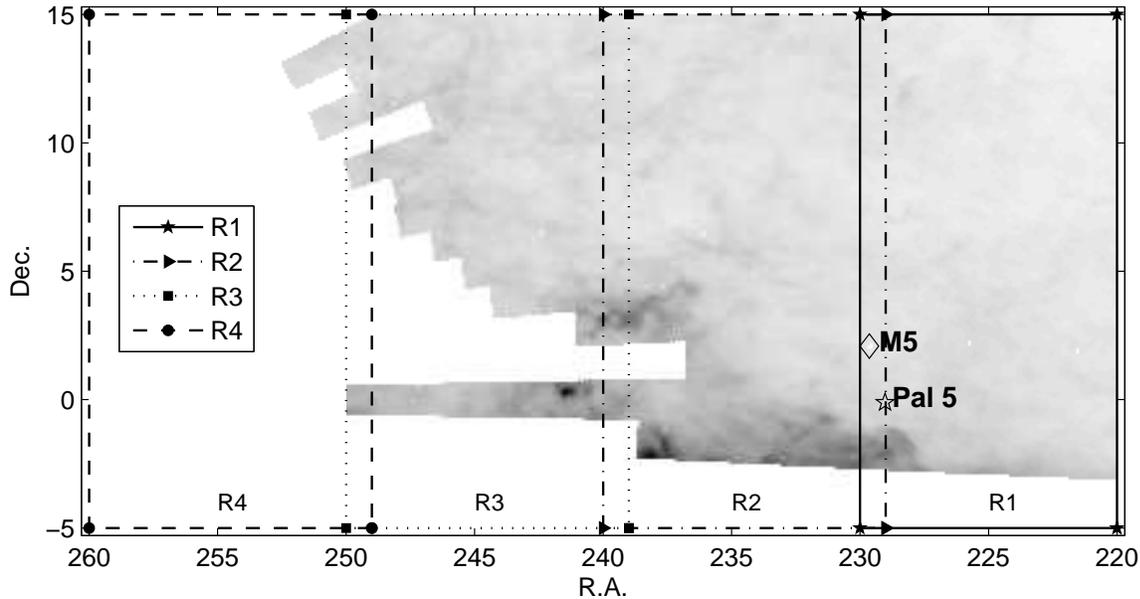}
    \caption{Distribution of interstellar extinction from the Milk
    Way. The resolution of this image is 6 arcmin, and E(B-V)s used
    by the SDSS databse are derived from \citet{sc98}. As
    mentioned previously, the blank regions are M5, bright stars and
    areas not observed by the SDSS.}
    \label{f2}
\end{figure}

Fig. \ref{f2} presents the interstellar extinction
distribution derived based on the values of $E(B-V)$ from
\citet{sc98}. The resolution of the distribution is about 6 arcmin.
There is smaller extinction in the northwest of the sky, but larger
extinction in the southeast on the whole. The reddening correction
in magnitude ($E(B-V)$) is 0.057 in average, and the maximum and
minimum is about 0.384 and 0.016, respectively. All the sources in
the sample are dereddened.

\begin{figure}
   \centering
   \includegraphics[width=12cm]{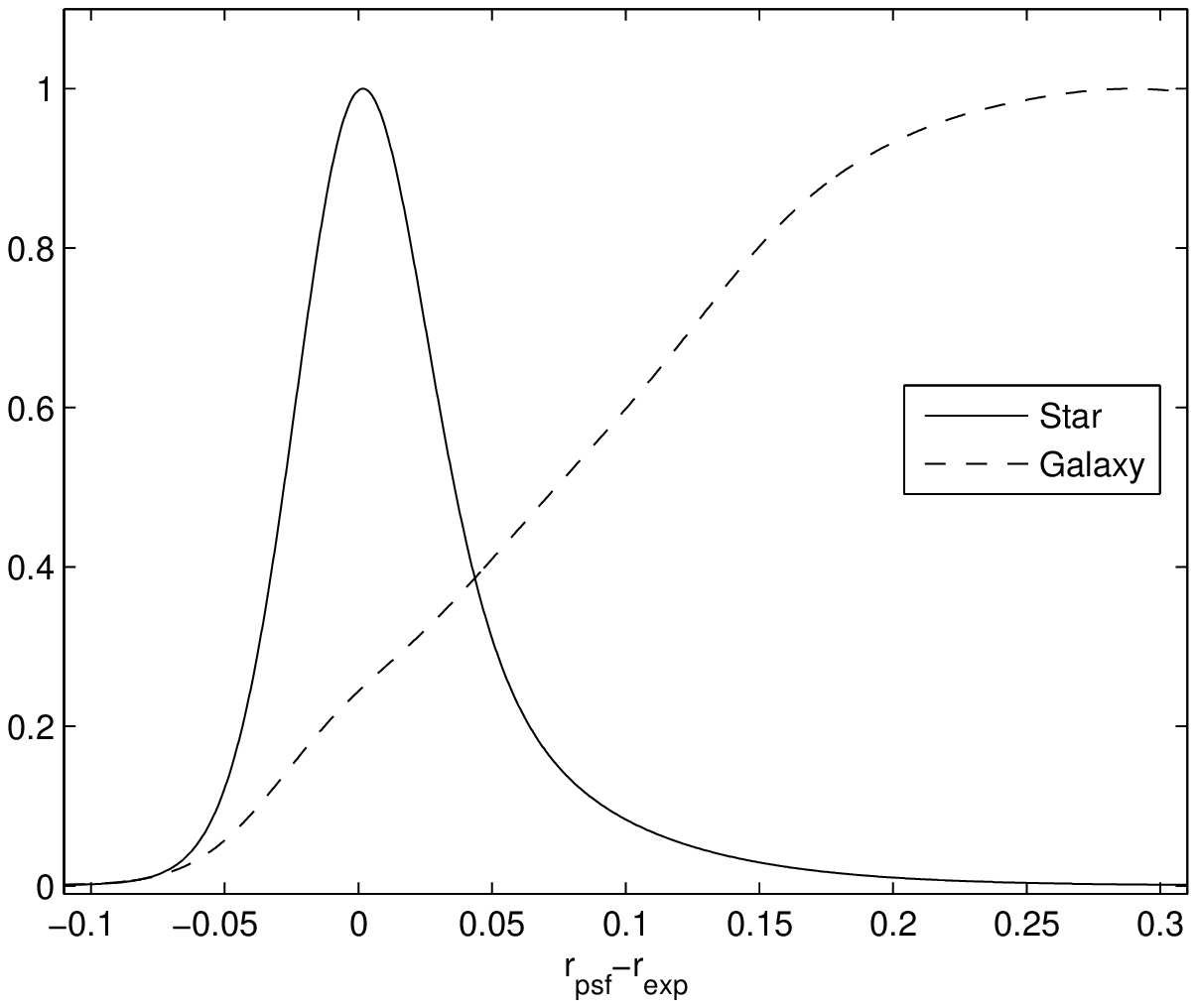} %[width=13cm]
   \caption{The distribution diagrams of $r_{\textrm{psf}} - r_{\textrm{exp}}$
   for both star and galaxy. Kernel density estimation is used
   to fit the distributions with normal kernel and 0.02 bandwidth\citep{bo97}.
   The two distributions are normalized to the peak value.}
   \label{f3}
\end{figure}

Following \citet{be06}, the difference between the magnitudes
obtained by PSF photometry and by fitting an exponential profile,
namely, $r_{\textrm{psf}}-r_{\textrm{exp}}$ distributions of both
star and galaxy are plotted in Fig. 3. It is clear that stars are
tightly concentrated around zero, whereas galaxies reveal a
significant positive excess. Except for the cases mentioned in the
previous section, classification is unauthentic when galaxies are
point-like. PSF is the best fit model to estimate the magnitude of a
point source ($r_{\textrm{psf}}$), while exponential model is one of
models to calculate the magnitude of an extended one
($r_{\textrm{exp}}$). SDSS constructs a simple classifier with the
analogous difference between $r_{\textrm{psf}}$ and
${r_{\textrm{exp}}}$. In Fig. \ref{f3}, $r_{\textrm{psf}}
- {r_{\textrm{exp}}}$ is concentrated on zero for point-like
sources, while it is far away from zero for extended sources. Both
of them have an intersection where stars and galaxies cannot be
distinguished. Also following \citet{be06}, we take an threshold
$r_{\textrm{psf}} - {r_{\textrm{exp}}} = 0.05$ as the division between stars and galaxies.
As a result, a sample of 4,082,662 point sources
remains, which includes 1,079,301 in R1, 1,478,788 in R2, 1,378,560
in R3 and 146,013 in R4.

\section{Back Propagation Neural Network} \label{BPNN}
BPNN is now applied to various areas including astronomy, such as
pattern classification, face and speech recognition and finance
\citep{hn96,hy98}. In astronomy, BPNN is used as classifier in both
photometric and spectral aspects \citep{hi94,fo96} or morphological
recognizer \citep{od92,na95} in imaging or value estimator
\citep{ba00} in determining theoretical models and physical
parameters. Virtually, the specifical mechanism of BPNN is as
follow: first, provide a train-test data set and train the
configured BPNN with them, just as a teacher teaches a student to
tell him which is a cluster star and which is a field one; then the
BPNN learns the knowledge again and again to modify its inner
configuration and makes itself perform well. During the course, BPNN
will be judged by test data with the learned prior knowledge; at
last, through this kind of repeated train-test-modify cycles, the
classifier (BPNN) gains the features and has the best-learned
experience to challenge new things. A data processing tool {\sc
matlab}, which provides a special neural network toolbox to design
and realize all kinds of ANNs, is introduced in our work. The
definition of neural networks, and other technical terms as well as
the specified process of BPNN are described in the appendix.

As mentioned previously, \citet{be06} used neural networks to
reconstruct the probability distribution of cluster stars with the
SDSS $ugriz$ photometric data. The idea of the approach is very
simple: with $ugriz$ 5-band photometric data of cluster members and
field stars as inputs of a BPNN, we get an estimation of the
probability of cluster member as the output after the network is
best trained. This method makes full use of photometric information
and constructs a probability estimator in high dimensional data
space with limited resources. When being trained, the BPNN can
pick out bad sources automatically to form an accurate separator.

With the sample picked out strictly, the first step to detect
the tidal tails is to figure out all the cluster members of
Palomar 5 in the selected field. A lot of pattern recognition
techniques are available, such as Bayes classifier, template
matching, clustering analysis and artificial neural networks (see,
e.g., Sergios \& Konstantinos 2006, and references therein).
In present study, the only thing we need to do is to estimate the membership
probability of an object. BPNN can measure the posterior probability
$P(C|\textbf{\emph{x}})$ in high dimension space, where $C$ denotes
the cluster member class and $\textbf{\emph{\textbf{x}}}$ is the
photometric data vector.

At last, we make a summary of the basic parameters and components of
BPNN used in our paper. First, the dimension of input layer is 5
(photometric magnitudes) , and the transfer function of each neuron
is Log-Sigmoid function. Then, the dimension of output vector is
one, which yields 0 (for field stars) or 1 (for cluster stars). Mean
squared error (MSE) is used to calculate the deviations between the
output of BPNN and the real object type value. Finally, the
Levenberg-Marquardt Backprogation (LMBP) algorithm is used to train
the network to minimize the performance function MSE. The initial
state (such as initial weights and biases, condition of termination,
parameters of LMBP algorithm) is given automatically by {\sc
matlab}.

\section{Tidal Tails Detection Based on BPNN}
\label{appl}
\subsection{Data Set Selection for Training and Test}
\label{data-cmds} In order to apply the BPNN method to the
tidal tails detection of Palomar 5, first of all, a training and test data set
should be chosen to guide the BPNN to learn the knowledge of the
cluster, so that it has the ability to figure out the probability of
cluster member. As \citet{be06} suggested, we select cluster stars
from candidates using color-magnitude diagrams.

\begin{figure}
   \centering
   \includegraphics[width=12cm]{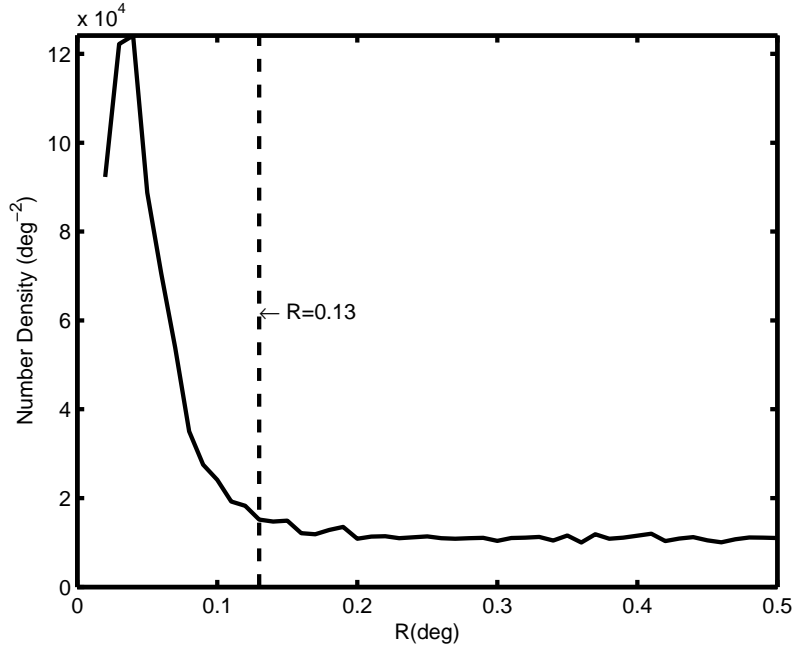} %[width=13cm]
    \caption{The radial number density distribution around the center of
    Palomar 5. The solid line shows that the density declines as the radius increases.
    The dashed line gives a threshold $R$ = 0.13$\dg$.
    Stars within the threshold radius are considered as cluster member candidates.}
    \label{f4}
\end{figure}

Along the main-sequence in the CMD, stars around the center of a
cluster within a proper radius are likely to be members. Fig.
\ref{f4} shows the radial number density distribution
of stars around Palomar 5. We can see that the density descends from the center
to the external of the cluster. As $R > 0.25\dg$, the average number
density becomes about $1.0958\pm0.0497\times10^{4}$ deg$^{-2}$.
Radius $R$ = $0.13\dg$, where the density is a little higher than
the average to avoid excessive field stars, is used to select
candidates of cluster stars. About 1523 cluster member candidates
are reserved.

\begin{figure}
  \centering
  \subfloat[]{
    \label{f5a}
    \includegraphics[width=2.8in]{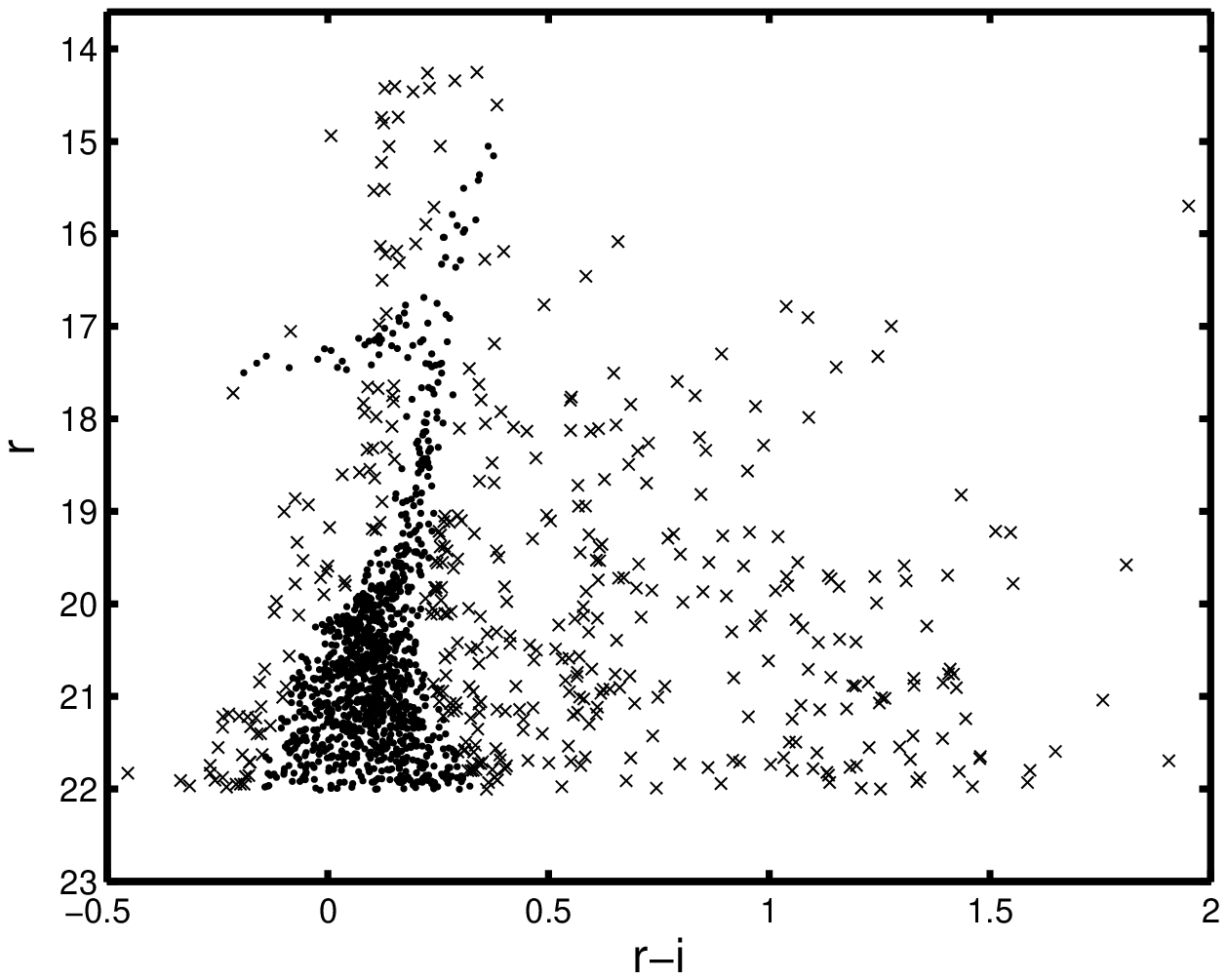}}
  \hspace{0.0cm}
  \subfloat[]{
  \label{f5b}
  \includegraphics[width=2.8in]{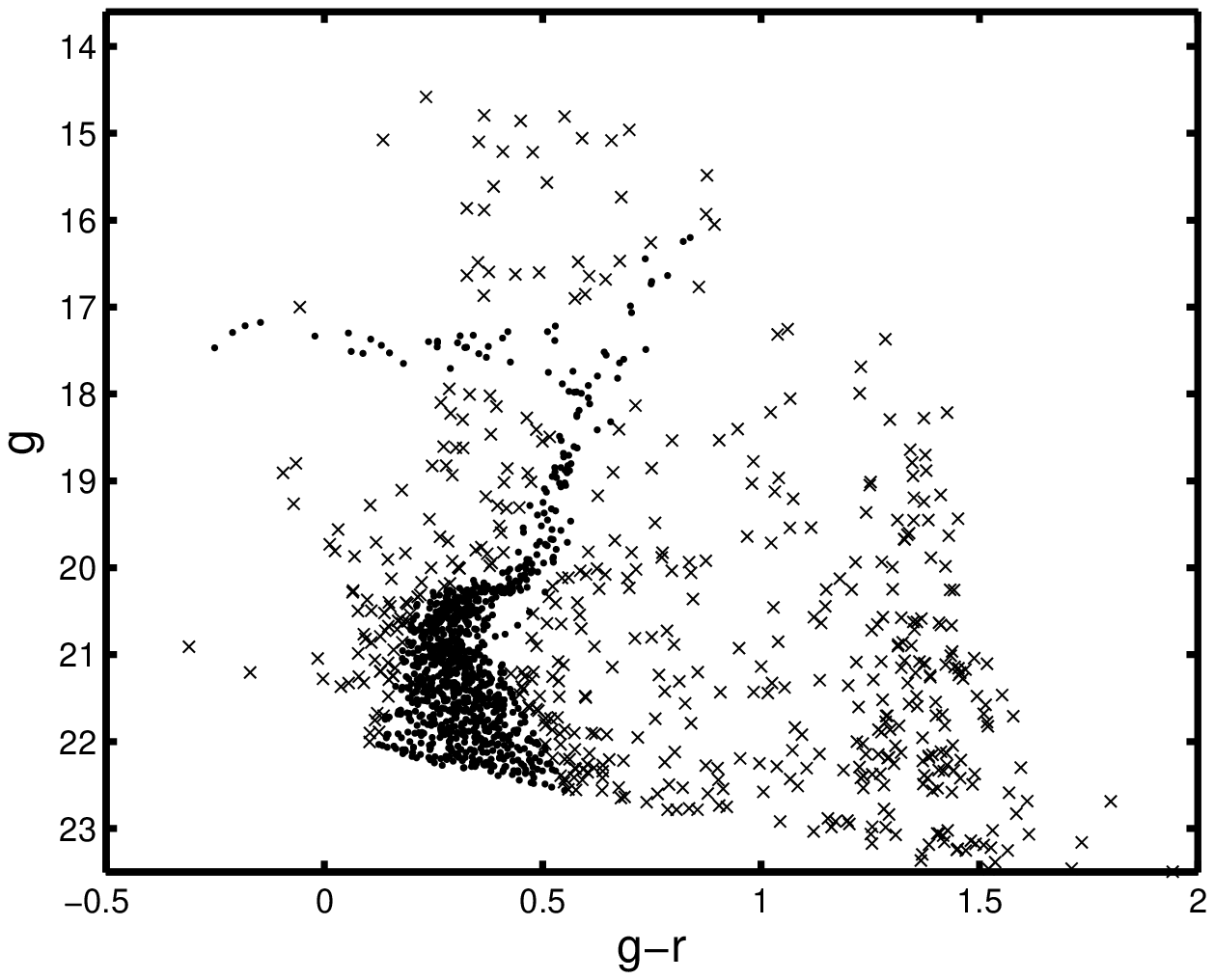}}
  \vspace{0.0cm}
  \subfloat[]{
  \label{f5c}
  \includegraphics[width=2.8in]{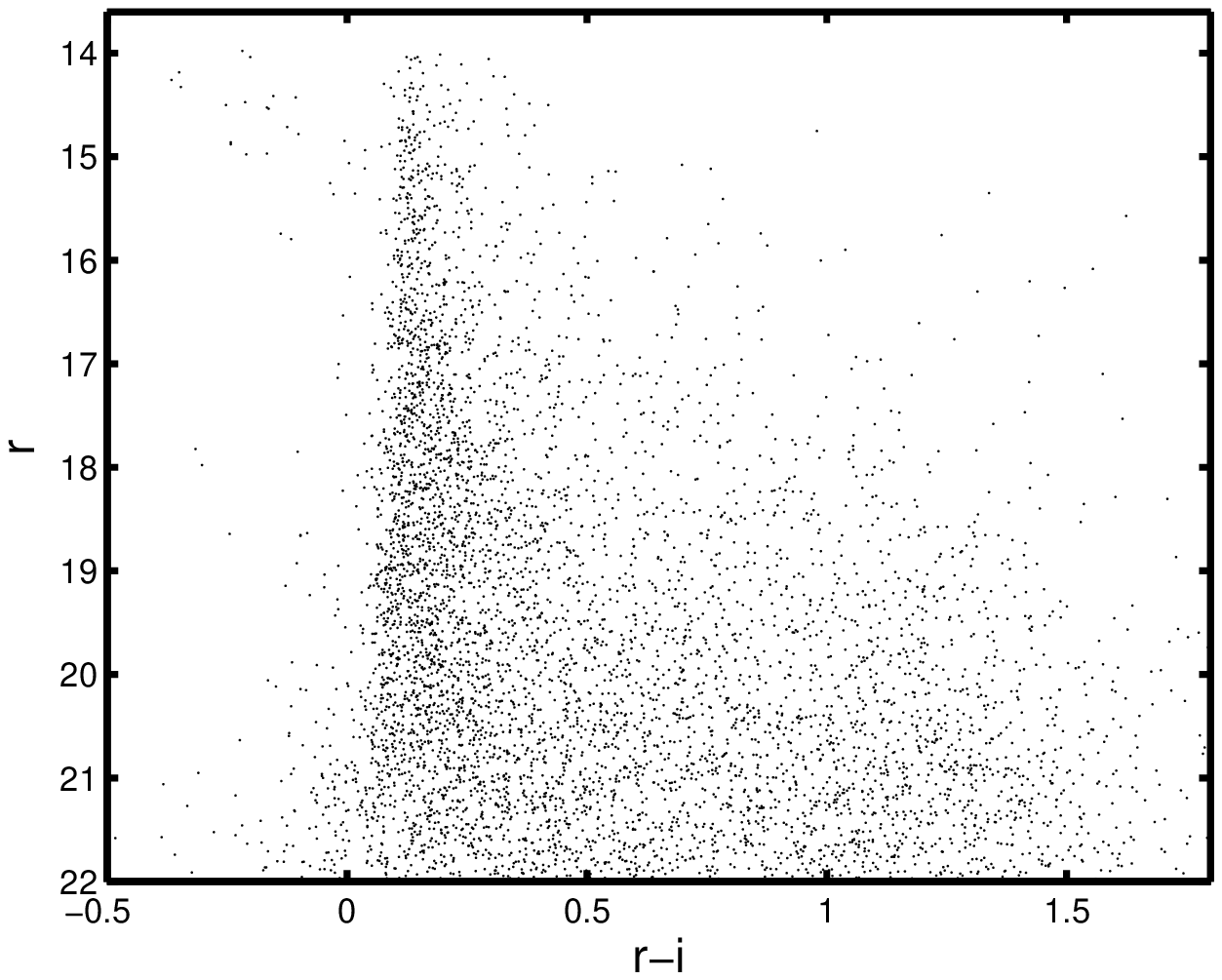}}
  \hspace{0.0cm}
  \subfloat[]{
  \label{f5d}
  \includegraphics[width=2.8in]{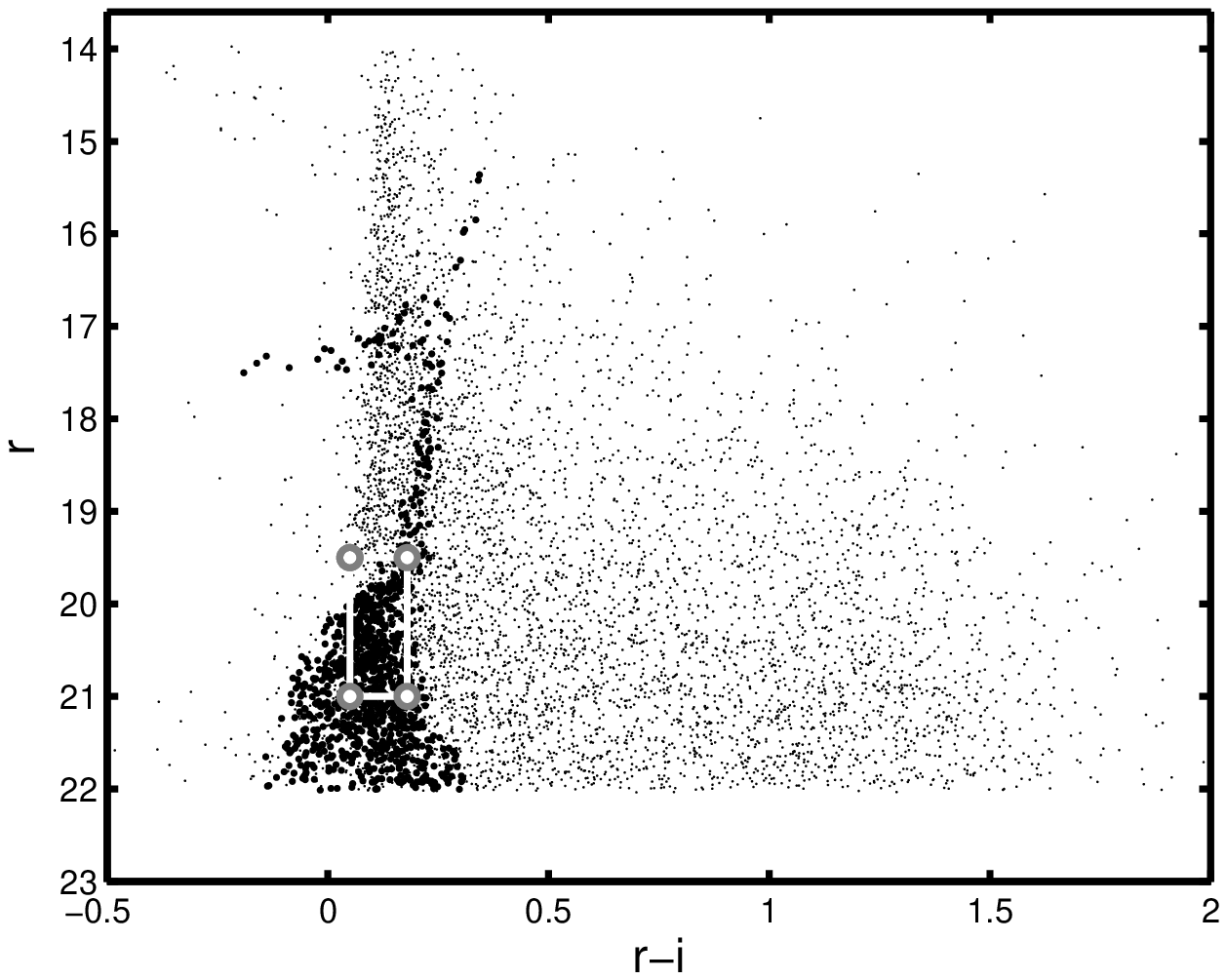}}
  \caption{The color-magnitude diagrams used to select cluster and field
  stars. a) The CMD of $r$ vs $r-i$. Bold dots are the possible cluster members
  and crossings denote the excluded non-cluster sources. b) The CMD of $g$ vs $g-r$,
  used to assist $r$ vs $r-i$ CMD to pick out cluster stars from candidates.
  c) The distribution of the selected field stars in $r$ vs $r-i$ CDM. d) The $r$ vs $r-i$ CMD,
  including both the extracted cluster members and field stars. The smaller dots are field stars and
  the larger dots are cluster members. The white box with four circles shows the enclosure to decide
  the number of field star samples.}  \label{f5}
\end{figure}

The next step is to pick out the most probable cluster stars from
the 1523 candidates. Fig. \ref{f5} demonstrates
the process of selecting cluster stars as a part of the training and
test data set. Objects are reserved by encircling the CMD of $r$
vs $r-i$ (Fig. \ref{f5a}) and $g$ vs $g-r$(Fig. \ref{f5b})
with proper enclosures. These objects
may be main-sequence stars, red giants and blue horizontal branch
stars belonging to Palomar 5. In this way, contaminations from
remanent galaxies and field stars (crosses in both CMDs) are almost
eliminated. By intersection of these two CMDs, 957 objects are kept
down to form the distribution of cluster stars in Fig.
{\ref{f5d}} (bigger black dots).

For field-star selection, $R > 1\dg$, which is far enough from
the center of Palomar 5, is adopted to choose candidates of field
stars. Because the number of candidates is tremendous, it's
impossible to preserve all of them as our data set. Ideally, we hope
that roughly equal numbers of objects should lie in both sides of
the boundary in the data space so that they can fully represent these two
classes. Accordingly, we chose field stars randomly in the sky, satisfying
roughly equal numbers of both field and cluster stars near the
main-sequence turnoff in the CMD. Thus, a box (a white rectangle with 4 hollow
circles at the vertexes in Fig. \ref{f5d})with $19.5
\leq r \leq 21$ and $ 0.05 \leq r-i \leq 0.18 $ is designed to form
a region encircling the turnoff. There are about 320 stars in the
box of the CMD for the cluster and field, respectively. As a result, 6991 field stars (smaller dots in
Fig. \ref{f5c}) and 957 cluster stars constitute the
whole train-test data set (Fig. \ref{f5d}).
As mentioned previously, these field stars with target value 0 and
cluster stars with value 1 are transmitted to train a BPNN. The BPNN
will configure itself to estimate the possibility of cluster member
for each object. Although the remained data set may contain
impurities more or less, they are negligible relative to the total group
sizes. And at the same time, BPNN will get rid of them automatically
through being trained again and again, which is one of the reasons
why we consider BPNN as our method.

\subsection{Network Structure Determination}
In order to normalize the input data, we subtract the mean
magnitudes of all data and then divide them by their standard
deviations. Furthermore, for the sake of determining the number of
layers and neurons in each layer, the star sample is segmented into
training and test data sets. Training set is used to train a BPNN,
while test set is used to measure its performance. Since cluster
stars are relatively scarce, all of them are placed into the
training set. Half field stars are laid aside stochastically into
training data set and the others are regarded as test set.

\begin{figure}
   \begin{center}
   \epsscale{1}{1}
   \plotone{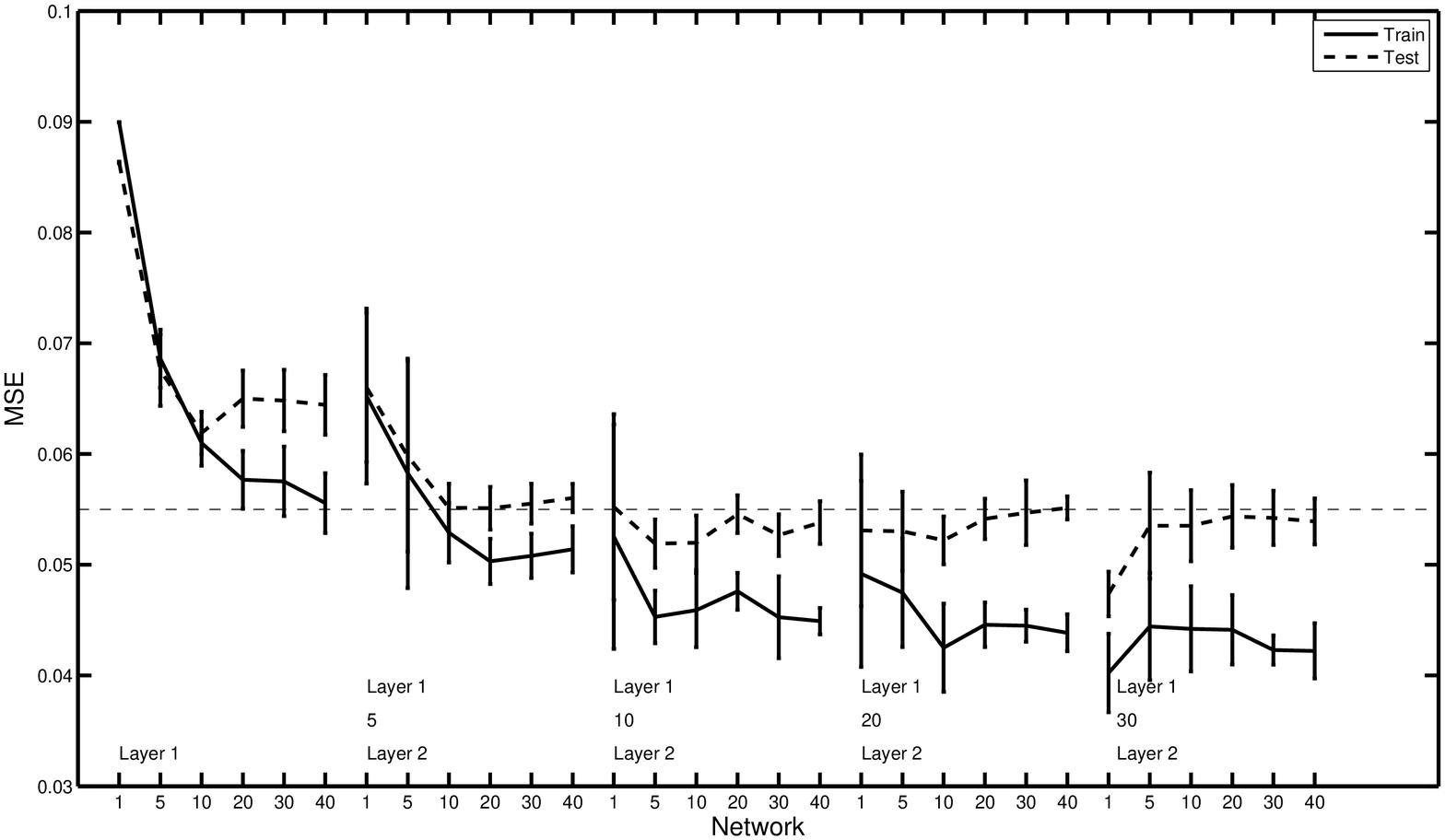}
   \caption{The mean training and test MSEs of various networks including
   1 or 2 hidden layers. The darker solid line shows the training MSEs,
   while the light dotted line shows the test MSEs. The dashed horizontal
   line (MSE = 0.055) presents one level so that it can help
   us to check the network with the best performance.}
   \label{f6}
   \end{center}
\end{figure}

Ten experiments are implemented for each designed BPNN to calculate
the average output as the probability of cluster member. In this
way, some random influences from initial weights, direction of
modifying weights and algorithm terminating conditions are weakened.
Here, one hidden layer and two hidden layers with 1, 10, 20, 30 and
40 neurons in the specified networks are investigated. The
performance of them is showed in Fig. \ref{f6}. Each network is
trained and tested ten times with data set gained in the previous
section. We stop the algorithm in every concerned network when the
test MSE reaches its minimum. As Fig.\ref{f6} indicates, with the
increasing complication of network configuration, the training and
test MSEs descend as a whole. And in the same network group (for
example Layer 1 = 10), the test MSE always descends at first and
then ascends when the number of neurons in the second hidden layer
increases. No larger difference among the BPNNs with 2 hidden layers
when the number of neurons in the first hidden layer is lager than
10. Considering about the test performance and the complexity of the
configuration relevant to the speed of training, Net[5:10,10,1],
which has 5 input elements, 10 neurons in the first hidden layer, 10
in the second hidden layer and 1 output, serves as our final model.

\subsection{Detect Tidal Tails With the Best Trained Network}
\begin{figure}
   \centering
   \includegraphics[width=12cm]{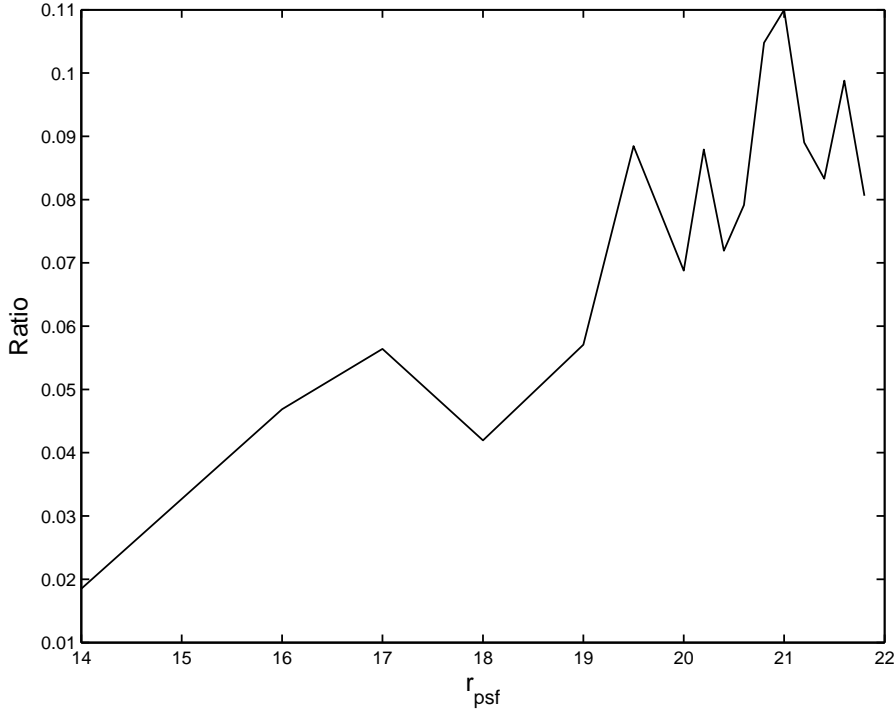} %[width=13cm]
   \caption{The relation between the ratio of mean outputs for field and
   cluster stars and the $r_{\textrm{psf}}$ magnitude.}
   \label{f7}
\end{figure}

It's necessary to take into account the effectiveness of the trained
networks. So, we construct a function of the magnitude
($r_{\textrm{psf}}$) as the ratio of the mean output for field
and cluster stars in the training and test data set. Ideally, the
ratio should be around zero as long as the networks are trained
well. In Fig. \ref{f7}, the ratio goes up gradually when
$r_{\textrm{psf}}$ becomes fainter. No significant deviation is shown above 0.1
near the magnitude limit. This indicates that the selection of the
training and test data set, and BPNNs are determined considerably
well. However, in order to trace out the profile of the tidal tails
more evidently, we cut off all the sources with a truncation
$r_{\textrm{psf}}$ = 21 with smaller photometric errors.

All the sources with normalized $ugriz$ magnitudes are imported into the
trained ten networks, and the mean probability of cluster member for
each star from the output layer is calculated. In order to present
the panorama of the distribution of cluster stars, we divide the
whole field into small square bins, whose sizes are $6'\times6'$.
About 90 objects are included in each bin, which is enough for
statistical analysis. Here, the mean probability in each bin is
computed. At the same time, Gaussian smoothing and median filtering are
used to get rid of noises when detecting the tails in the field. In
addition, in order to enhance the resolution of the distribution,
cubic spline interpolation is also employed.

A lot of experiments are implemented to investigate the factors
impacting the distribution. These tests include the parameters used
in smoothing tools, the selection of cluster stars and the field
stars chosen in different regions of the field. As a result, cluster
stars within the radius $R < 0.13\dg$ are appropriate for training the
networks. Larger $R$ will bring in pollution from field stars, while
smaller $R$ yields less cluster stars which cannot provide enough
information about Palomar 5 members. Thus, the selection criterion of
cluster stellar candidates in $\S$ \ref{data-cmds} is reasonable.
Field stellar samples chosen from different regions  make no difference
unless they cover the possible position of tidal tails. However, it
does not affect the detection of the rest parts.

\begin{figure}
   \begin{center}
   \epsscale{1}{1}
   \plotone{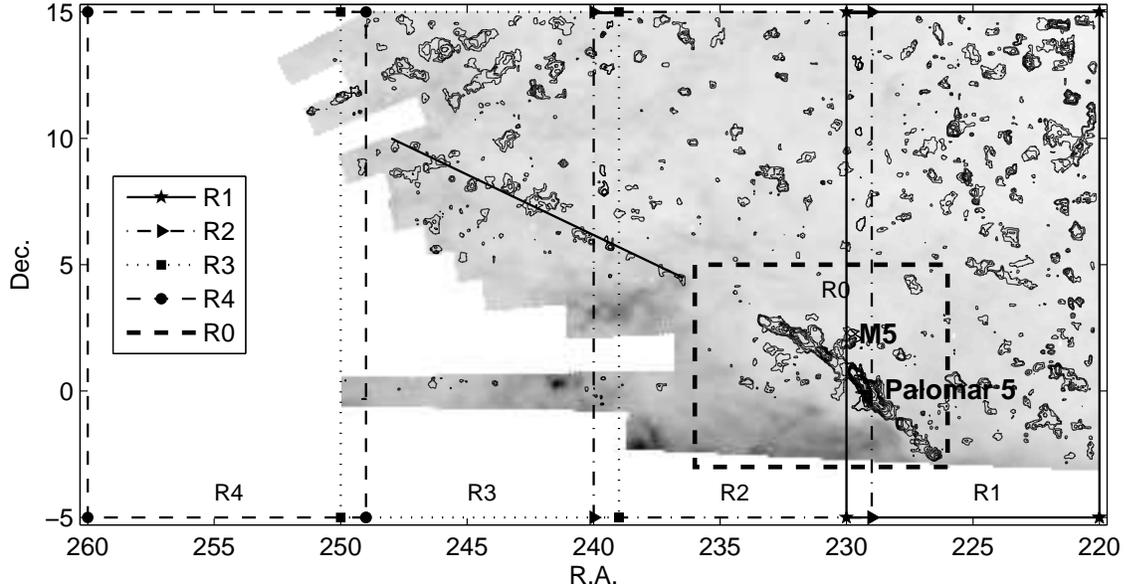}
   \caption{The contours of the smoothed probability distribution of cluster member
   with $1\sigma$ high above the mean. The solid line shows the possible extension, and dashed rectangle R0
   is the target region we will discuss. The overlapped grey background is the reddening distribution of
   the Galaxy.}
   \label{f8}
   \end{center}
\end{figure}

\begin{figure}
   \begin{center}
   \epsscale{1}{1}
   \plotone{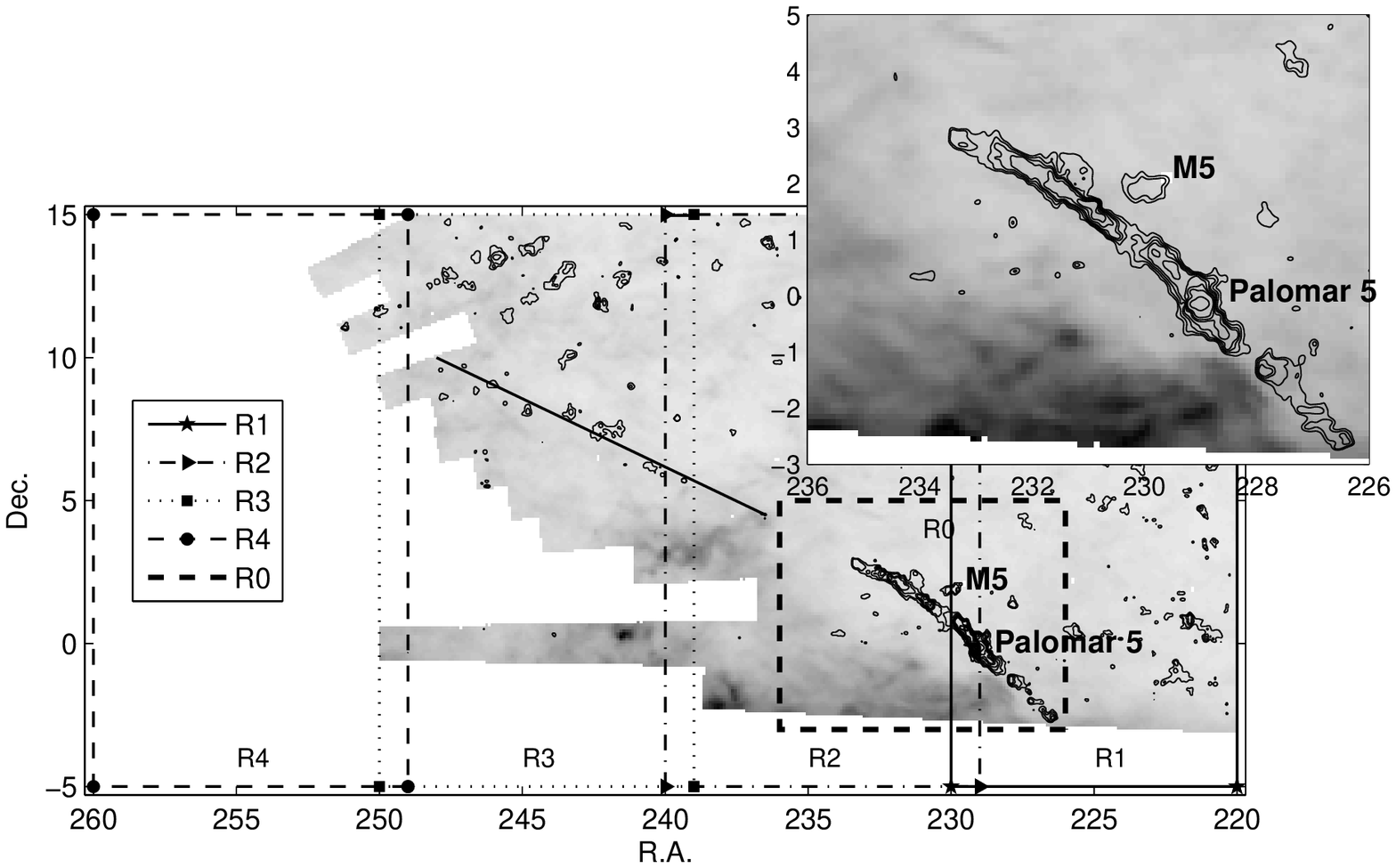}
   \caption{The contours of the smoothed probability distribution of cluster member with $1.5\sigma$ higher
   above the mean. The upper-right plot shows the recomputed distribution in the region R0. Detailed
   specifications should be referred to the text.}
   \label{f9}
   \end{center}
\end{figure}

Fig. \ref{f8} exhibits the contours of 2-dimension
probability distribution, where four regions are processed by the
same best trained network and smoothing techniques. Although
smoothed, the distribution is fluctuant in the whole field. We find
the $P(C)s$, the probability of cluster member in all bins, obey a
Gaussian distribution with mean value 0.0488 and standard
deviation $\sigma$ = 0.0087. In order to check the tails more widely
and to find possible longer tails, the contour levels are larger
than $1\sigma$ above the mean. In this figure, M5 is detected at the
same time, due to the similarity of their color-magnitude diagrams.
The black solid line traveling through R2 and R3 provides us with a
possible tidal tail away from the main tails in R0 ($226\dg < \alpha
< 236\dg$ and $-3\dg < \delta< 5\dg$). We are not sure that the tail
along this line is the real extension of the tails of Palomar 5,
because the foreground noises in the residual regions are so similar
to it. Fig. \ref{f9} shows the contours of the
probability distribution, where the contour levels are higher than
$1.5\sigma$ above the mean. The possible extension vanishes, but
some debris still coexist with foreground noises. However,
\citet{gr06b} insists that the extension is the real tail.
Therefore, only the region R0, which encloses the clear tidal tails
of Palomar 5, is considered as our region of investigation in the
next discussion. The sub-figure in Fig. \ref{f9}
shows the smoothed probability distribution with $1.5\sigma$ in R0.
There are some ignorable differences between the two plots.
These petty changes are caused by different bin coordinates and by region smoothing.

\section{Properties of The Tidal Tails and The Cluster}\label{diss}
\subsection{The Profiles of the Tails}
The subplot in Fig. \ref{f9} presents the holistic
smoothed distribution of Palomar 5 members. The whole tail lies away
from the dense regions of the Galactic extinction, which implies
that the extinction has little effect on our detection. Two tails
extend from the core region of Palomar 5 to southwest and northeast
directions, which we called South Tail (ST) and North Tail (NT),
respectively. ST is the leading part facing toward the Galactic
disk, while NT is the trailing part. The angular distances are
$5.42\dg$ for NT and $3.77\dg$ for ST. NT is a little shorter than
the $5.8\dg$ northern tail detected by \citet{od03}, while the lengths of both ST
are approximately equal. One reason may be that the smoothing flatten the distribution. In
addition, the longer debris along the north tail detected by \citet{od03} may be
the disturbance or noise of field stars. However, the main tails we
find are very similar to the results of \citet{od03}, while we do
not support the discovery of any longer tails to the north of the
sky that \citet{gr06b} reported to be a 22$\dg$ tail of Palomar 5,
because the more extensive debris is so semblable to the background
noises. There is no photometric data in the south, so we cannot
cover larger southern regions to make sure a longer ST.

Another fact we can see from the contour map in Fig. \ref{f9} is that the
distribution of cluster members is more dense in NT than that in ST.
The possible reason causing this situation may be that: the orbit
direction of NT is nearer to us than ST; given that the components
of both tails are the same, because of their large extension, the
average magnitude of NT ought to be brighter than that in ST. So,
due to detecting limit, quite a few cluster members in ST cannot be
observed or excluded by data preprocessing, although they are
detected. However, when investigating all the detected members in
both NT and ST, we do not find any large magnitude shift by
comparing the mean magnitudes of the two tails in any places having
the same distance from the cluster center. The maximum magnitude
difference (between the northeast and southwest ends) is about 0.1 mag
(NT is a little brighter than ST). Qualitative analysis indicates
that even if the inclination angle of the tails is large, the
magnitude difference is small when considering a specified star
lying in different position in the tails. The above fact tells that
it may be true that the track of NT is nearer to us, but it cannot
change the distinct density difference of both tails obviously.
Therefore, this kind of situation should be referred to some
dynamical processes, which will be studied in future.

\begin{figure}
  \centering
  \subfloat[]{
    \label{f10a}
    \includegraphics[width=2.7in]{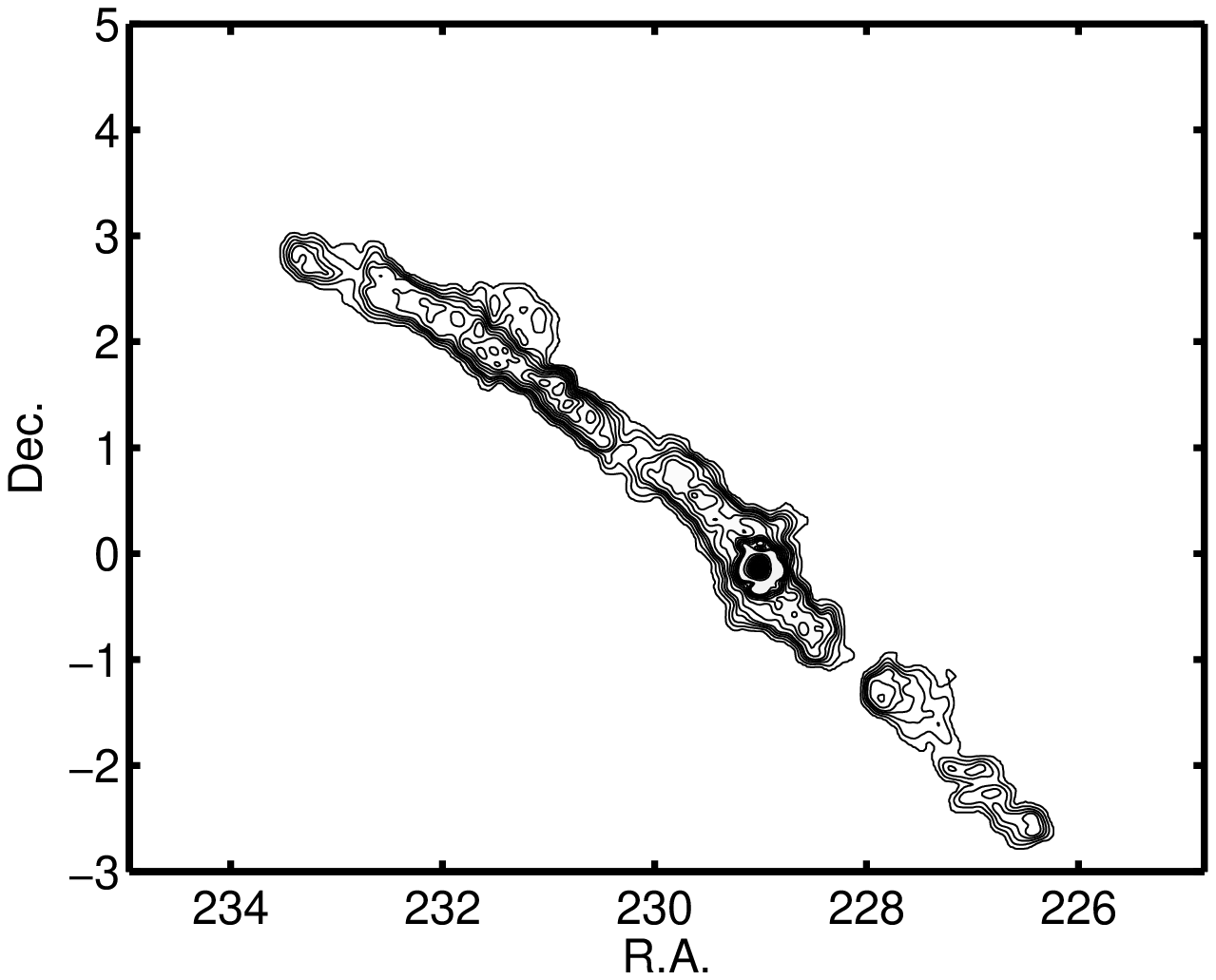}}
  \hspace{0.0cm}
  \subfloat[]{
  \label{f10b}
  \includegraphics[width=2.7in]{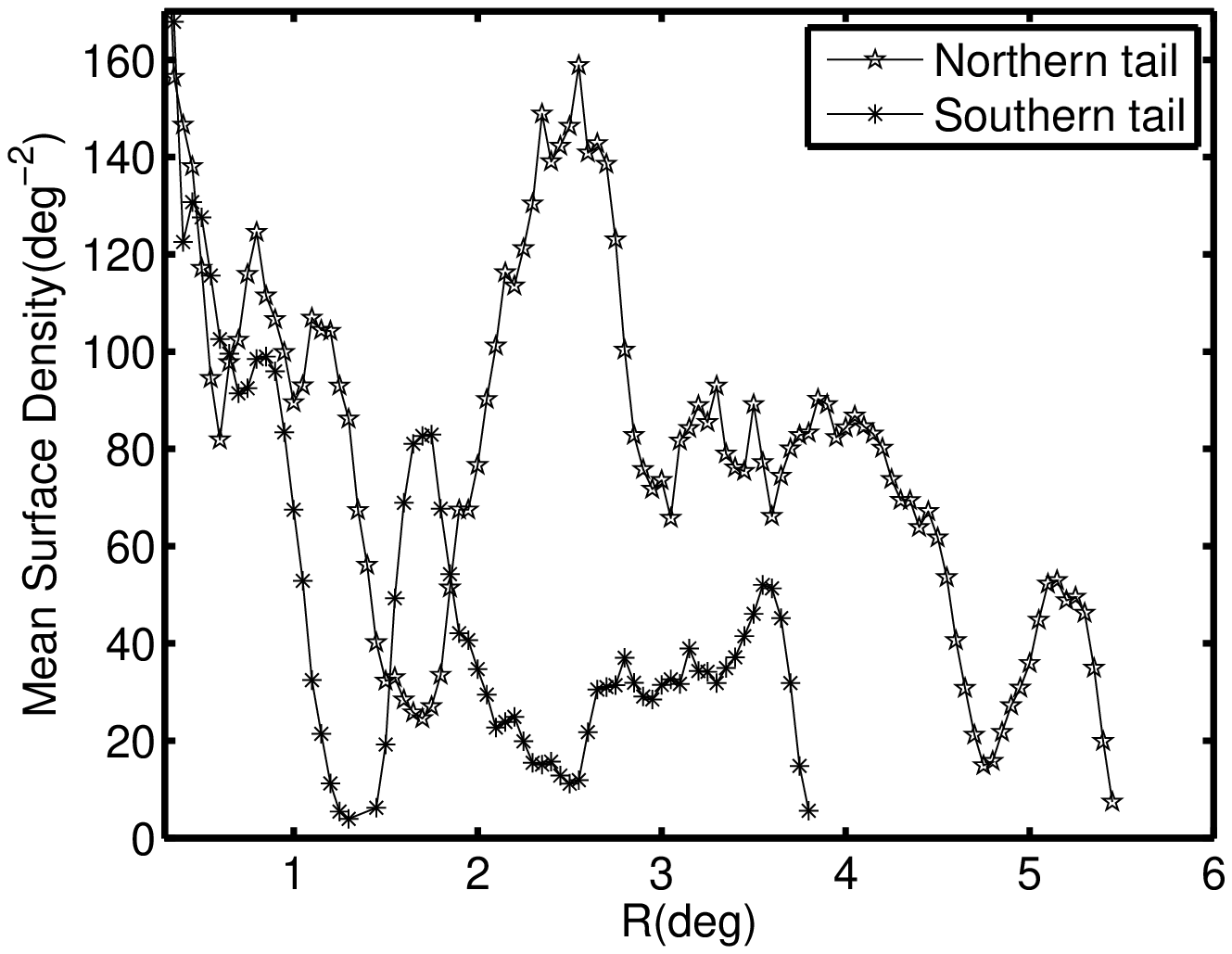}}
   \caption{(a) The surface density contour map of Palomar 5, converted from
   the smoothed probability distribution map. Here, M5 and noises around the tails
   are erased. (b) The radial profiles of surface density (deg$^{-2}$) for both tails.
   The interval of the radial distance is 0.05\dg.}
\end{figure}

We convert the probability distribution to surface density
distribution. First, we count the numbers of stars in all square bins. Then, the areas of all bins are
calculated. In this way, with the smoothed probabilities of cluster
member, we get the smoothed surface density distribution. Fig.
\ref{f10a} gives the transformed surface
density contour map which is interpolated by the technique of cubic
spline interpolation. Fig. \ref{f10a} indicates that there are no obvious changes compared with the
probability distribution contours. Furthermore, the noises around
the tails and M5 are gotten rid of from the map. From Fig. \ref{f10a},
we can see that there is no geometrical symmetry between ST and NT, and it
takes on an S shape near the center of Palomar 5. In fact,
\citet{de04} has illustrated this structure in their simulations.
Fig. \ref{f10b} shows the radial surface density
profiles of both ST and NT. In Fig. 10b, the numerical values originate
from the smoothed density. Thus the density is lower than the density
profile along the tails discussed by \citet{od03}.
Both of density profiles drop down quickly from the center of
Palomar 5. However, it seems to be that the density of NT descends
not so fast than the ST. In addition, the trailing tail seems to lag
behind the leading tail, because the density of ST reaches its peak
at $R$ = $1.7\dg$, while NT achieves its minimum. This phenomena can be
slightly seen in \citet{od03} when he discussed the radial profiles
of both tails, although he did not mention it. We cannot explain
what causes this phenomena yet, but it is an interesting result.
Possibly, this kind of lag is closely relative to the evolution of
the cluster and the interaction between Palomar 5 and the Galaxy.
Moreover, several stellar clumps emerge in both tails to form
substructures of the cluster. Some of them lie at around $R =
0.90\dg$ ($229.62\dg$, $0.56\dg$), $2.57\dg$ ($230.94\dg$,
$1.6\dg$), $3.28\dg$ ($231.56\dg$, $19.6\dg$) and $5.13\dg$
($233.28\dg$, $2.76\dg$) in the north, and $1.72\dg$ ($227.8\dg$,
$-1.32\dg$) and $3.50\dg$ ($226.5\dg$, $-2.54\dg$) in the south.
Maybe this kind of substructures and the radial density profiles are
formed by dynamical processes between the Palomar 5 and the Galaxy.
One possible explanation may be as follow: Palomar 5 has experienced
several encounters with the Galactic disk or bulge since its birth;
its body was heated by dynamical shocks; then the cluster stars were
accelerated, and gradually disrupted and extended along the moving
direction; as a result, some stars with small mass might escape from
the tails, and the residual ones constitute the substructures and
waited for the next shock. In fact, \citet{od03} has indicated that
Palomar 5 will be totally destroyed after the next disk crossing
within about 100 Myr.

\subsection{King Model Fitting and the Luminosity Function}
\begin{figure}
   \centering
   \includegraphics[width=12cm]{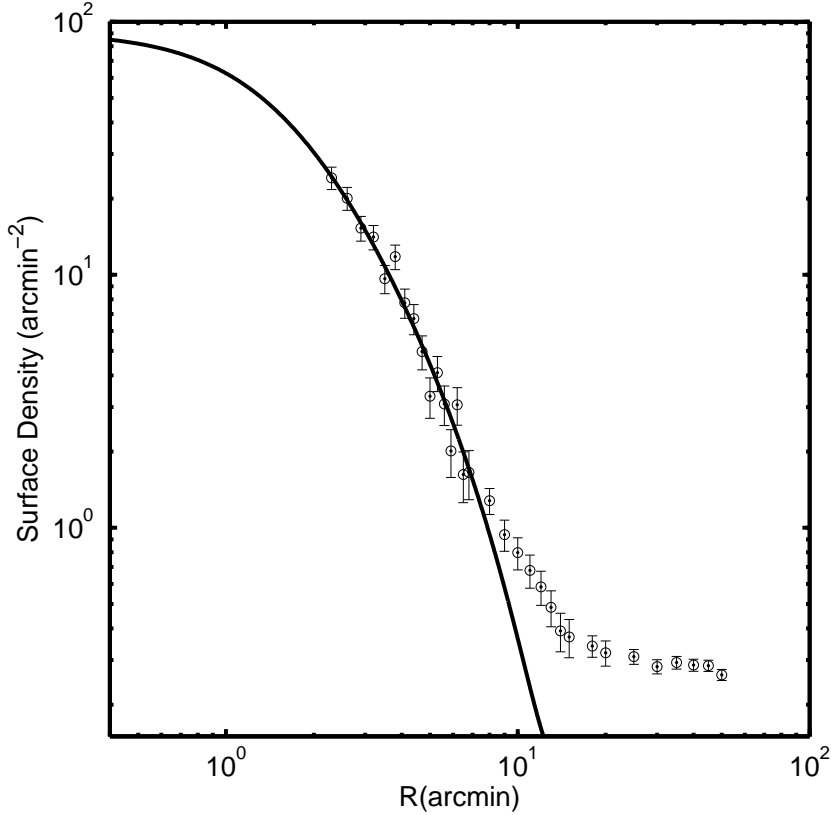} %[width=13cm]
   \caption{The radial surface distribution near the center of Palomar 5.  The circles represent
   the surface densities in all bins figured out by counting the cluster stars.
   The King model with the mean background number density (0.11) is drawn in the solid line.
   The coordinate system is a logarithmic system and the unit of density is arcmin$^{-2}$.
   }
   \label{f11}
\end{figure}

For density profile near the center, due to the
small size of Palomar 5, we take $R < 50'$ as the range for
discussion. Cluster stars are counted in all bins (from $2'$ to
$6.8'$, the bin size is $0.3'$; from $8'$ to $15'$, the bin size is
$1'$; the bin size of rest is $5'$), and the area of each annulus is
calculated. Fig. \ref{f11} shows the radial
surface density profile, and the best fitting of King model \citep{ki62}. King
model is expressed as:
\begin{equation}
\rho =
k\left\{\frac{1}{\left[1+(R/r_c)^2\right]^{\frac{1}{2}}}-\frac{1}{\left[1+(r_t/r_c)^2\right]^{\frac{1}{2}}}\right\}^2,
\end{equation}
where $k$ is a constant, $r_c$ is the core radius, $r_t$ is the
tidal radius and $R$ is the distance away from the center of the
cluster. Thus, the centric density is
\begin{equation}
\rho_0 = k\left\{1-\frac{1}{\left[1+(r_t/r_c)^2\right
]^{\frac{1}{2}}}\right\}^2.
\end{equation}
$c = r_t/r_c$ is the concentration ratio. We fit the data
subtracted from the background density (about 0.11) and obtain $k =
114.9$, $r_c = 1.60'$, $r_t = 14.29'$. Therefore, the concentration
ratio $c$ is 8.95 and $\rho_0$ is 90.81 arcmin$^{-2}$. We find that
the radii obtained in this paper are smaller than those of
\citet{ha96}: $r_c = 3.25'$ and $r_t = 16.28'$. From Fig. 11, we can
see clearly that the King model cannot fit the observed profile at
the outer regions ($R > 8'$) because of the tidal tails.

\begin{figure}
  \centering
  \subfloat[]{
    \label{f12a}
    \includegraphics[width=2.7in]{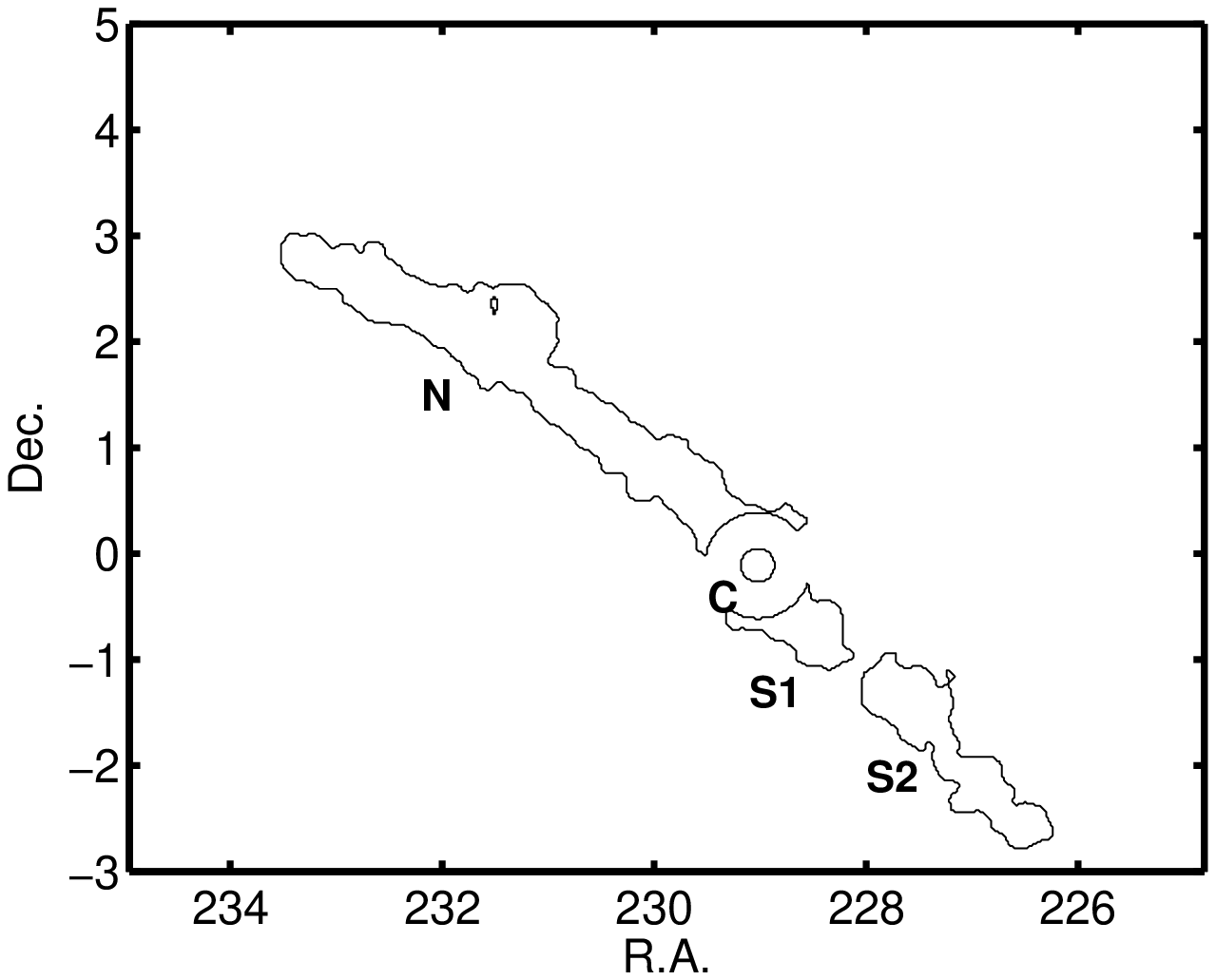}}
  \hspace{0.0cm}
  \centering
  \subfloat[]{
  \label{f12b}
  \includegraphics[width=2.7in]{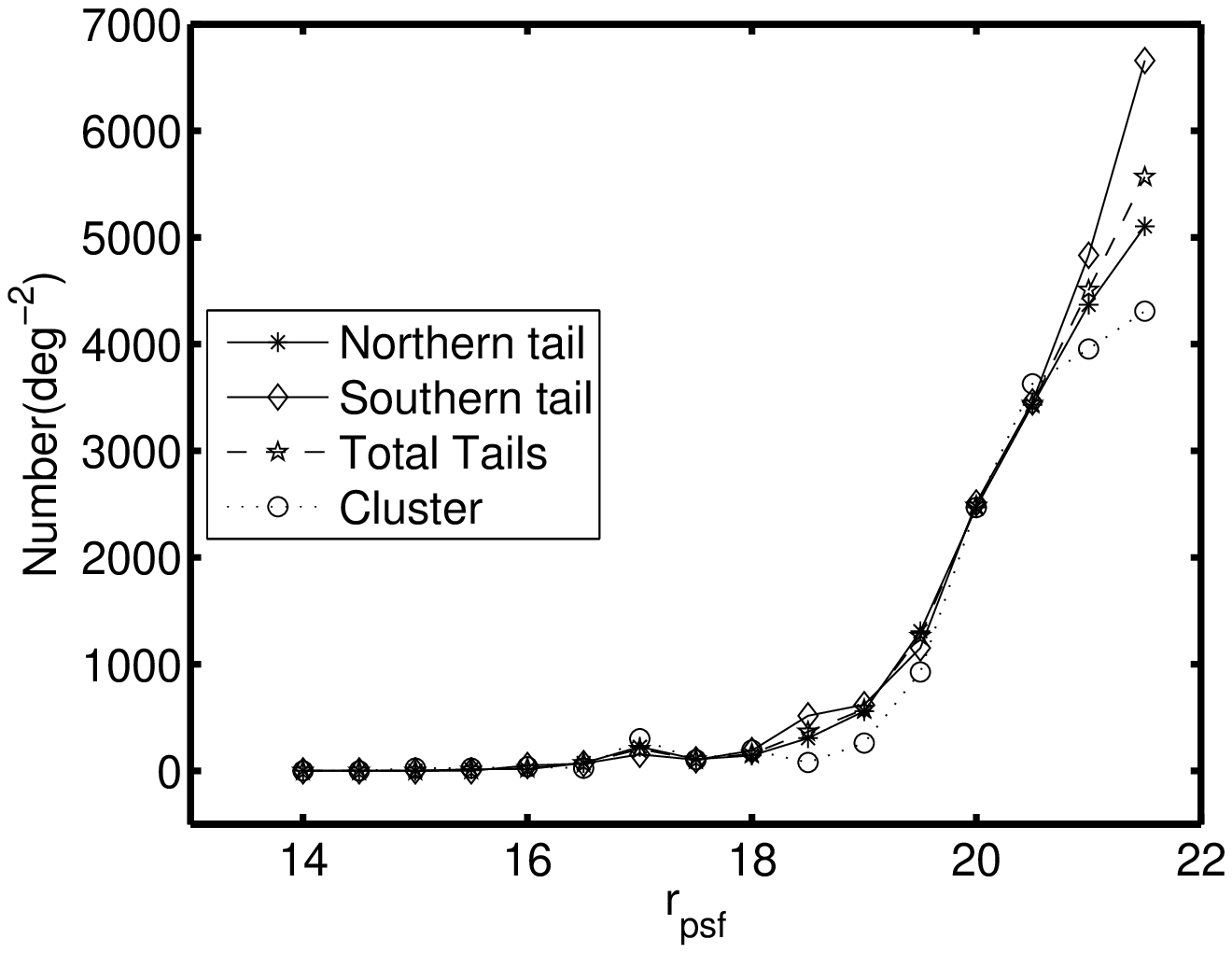}}
   \caption{(a) The boundaries used to calculate the LFs. C is the boundary of cluster itself
    with $R < 10'$; N is the boundary of the north tail; S1+S2 is the boundary of the south tail.
    The boundaries of both NT and ST are far away from the center with $R > 0.5\dg$. (b) Different
    luminosity functions. All LFs are rescaled by some factors to match the
    LF of the cluster.}
\end{figure}

Luminosity function (LF) of the cluster reflects its mass
distribution. We examine whether the luminosity functions of the
cluster itself and the tidal tails are the same, and try to
verify that the tidal tails come from Palomar 5. We examine
the luminosity functions in three regions: the north tail (N), the
south tail (S1 $\&$ S2) and the Palomar 5 (C). Fig.
\ref{f12a} shows the boundaries of these regions.
Radius $R = 10'$ is adopted for Palomar 5, and is near enough to the
center of the cluster to eliminate the contamination from the tails.
Radius $R > 0.5\dg$, far enough away from the center, separates the
tails into N for the north tail and S1 $\&$ S2 for the south tail.
Star counts are taken to deduce the LFs of the ST and NT in these
regions. We only consider the stars with high probability of cluster
member ($> 0.5$), which subsequently subtracts the contamination of
foreground field stars. Consequently, in Fig. \ref{f12b},
there are 4 LFs of ST, NT, total tails and the cluster itself, and
all of them are rescaled to match the LF of the
cluster. In the magnitude range of $19.0 \leq r_{\textrm{psf}} \leq 20.0$, LFs
are rescaled by factors 19.41, 22.43, 20.35 for NT, ST and total
tails, respectively. On the whole, there is little difference among
the LFs when $r_{psf} < 20.5$. As $r_{psf} \geq 21$, the LF
of the cluster lies lower than those of tails. This case confirms
so-called `mass segregation effect' \citep{ko04}, which shows that
cluster members with big mass will accumulate near the center
because of the loss of kinetic energy when colliding with others,
while stars with small mass would escape from the cluster into its
tails. Thus, these luminosity functions reveal the fact that the
stars in the tails come from Palomar 5, and some relevant physical
properties did not change much in its history.

\section{conclusion}\label{conc}
In this paper, we present a new method, Back Propagation Neural
Network, to detect the tidal tails of globular cluster Palomar 5.
Although some approaches such as matched-filter method \citep{ro02}
are widely applied to identifying the tails, we choose BPNN as our
model to find the exact and distinct tails of Palomar 5. The
photometric magnitudes of 5 bands ($ugriz$) in the SDSS DR6 are the
unique inputs and consequently the probability of cluster member is
the output in a best trained BPNN. BPNN resembles a black box, and
we need not consider its detailed inner structure. The only thing
we should do is to give it a set of well-selected cluster and field
stars (they may not be completely accurate) as a teacher to make
BPNN learn the knowledge. After gaining information, BPNN can
estimate the probability of the cluster member.

First of all, we obtain about 15,305,060 objects in a $40\times20$
deg$^2$ field ( $220\dg \leq \alpha \leq 260\dg$ and $-5\dg \leq
\delta \leq 15\dg$). Considering the effectiveness of star/galaxy
classification in the SDSS and the completeness of the observation,
we leave behind about 4,082,662 point sources (stars) with $14 \leq
r_{\textrm{psf}} \leq 22$ after reddening correction and eliminating
the pollution from galaxies as much as possible with the help of the distribution
map of $r_{\textrm{psf}}-r_{\textrm{exp}}$. Next, we make use of
surface density and CMDs to extract cluster stars from candidates,
which lie in the circle where $R < 0.13\dg$. And field stars
used to be trained are chosen by making equal numbers of both
cluster and field stars around the turnoff of main-sequence of
Palomar 5. In this way, about 960 cluster stars and 6800 field stars
are kept aside as the training and test data set to provide their
inherent characteristic information for BPNN. With the training and
test data, the best parameters and structures of BPNN are determined.
Then the best trained BPNN with 5 nodes in the input layer, 10
neurons in the first hidden layer, 10 neurons in the second hidden
layer and 1 neuron as output, is gained to compute the probability
estimation of cluster member for each point
source. We divide the field into bins with size $6'\times6'$ to
calculate the mean probability distribution. The impact of the
selection of field stars is also investigated and their effect is
not important.

S-shape tidal tails are detected, which subtend towards northeast
and southwest from the center of the cluster: the trailing tail and
the leading tail, respectively. The angular distances are $5.42\dg$
for the north tail and $3.77\dg$ for the south one. At the same
time, there are some density clumps as the substructures of Palomar
5 in both tails. We cannot find any longer stretch for the NT if we
do not regard the extension far away as tails, and we cannot confirm
whether the ST has a longer spread because no photometric data are
available outside to the southwest. We also find an interesting
phenomenon from the radial profile of the density: the NT seems to
lag behind the ST like wave propagation, which may be caused by the
tidal shocks when Palomar 5 crossed the Galactic disk or bugle in
its history. In addition, we fit the radial density profile near the
cluster center with the King model and find that the model can fit
this kind of remote globular clusters with low density very well
when the radial distance is less than $8'$. However, when the radial
distance becomes larger, the density drops more slowly due to the
tidal tails. The tidal radius obtained in this paper is $14.29'$ and
the core radius is $1.68'$. Luminosity functions of both tails and
the cluster are also determined. We find that there is little
difference among the LFs of both tails coming from the original
cluster, and their properties have not changed evidently during
their lives.

\begin{acknowledgements}
We are grateful to the referee for thoughtful comments and
insightful suggestions that improved this paper greatly. This work
has been supported in part by the National Natural Science
Foundation of China, No. 10633020, 10603006, 10778720, 10873016, and
10803007; and by National Basic Research Program of China (973
Program) No. 2007CB815403. Z.-Y. W. acknowledges support from the
Knowledge Innovation Program of the Chinese Academy of Sciences.

Funding for the Sloan Digital Sky Survey (SDSS) and SDSS-II has been
provided by the Alfred P. Sloan Foundation, the Participating
Institutions, the National Science Foundation, the U.S. Department
of Energy, the National Aeronautics and Space Administration, the
Japanese Monbukagakusho, and the Max Planck Society, and the Higher
Education Funding Council for England. The SDSS Web site is
http://www.sdss.org/.
\end{acknowledgements}

\appendix                %%appendicial material is supported
\section{The Mechanism of BPNN}
\begin{figure}
   \plottwo{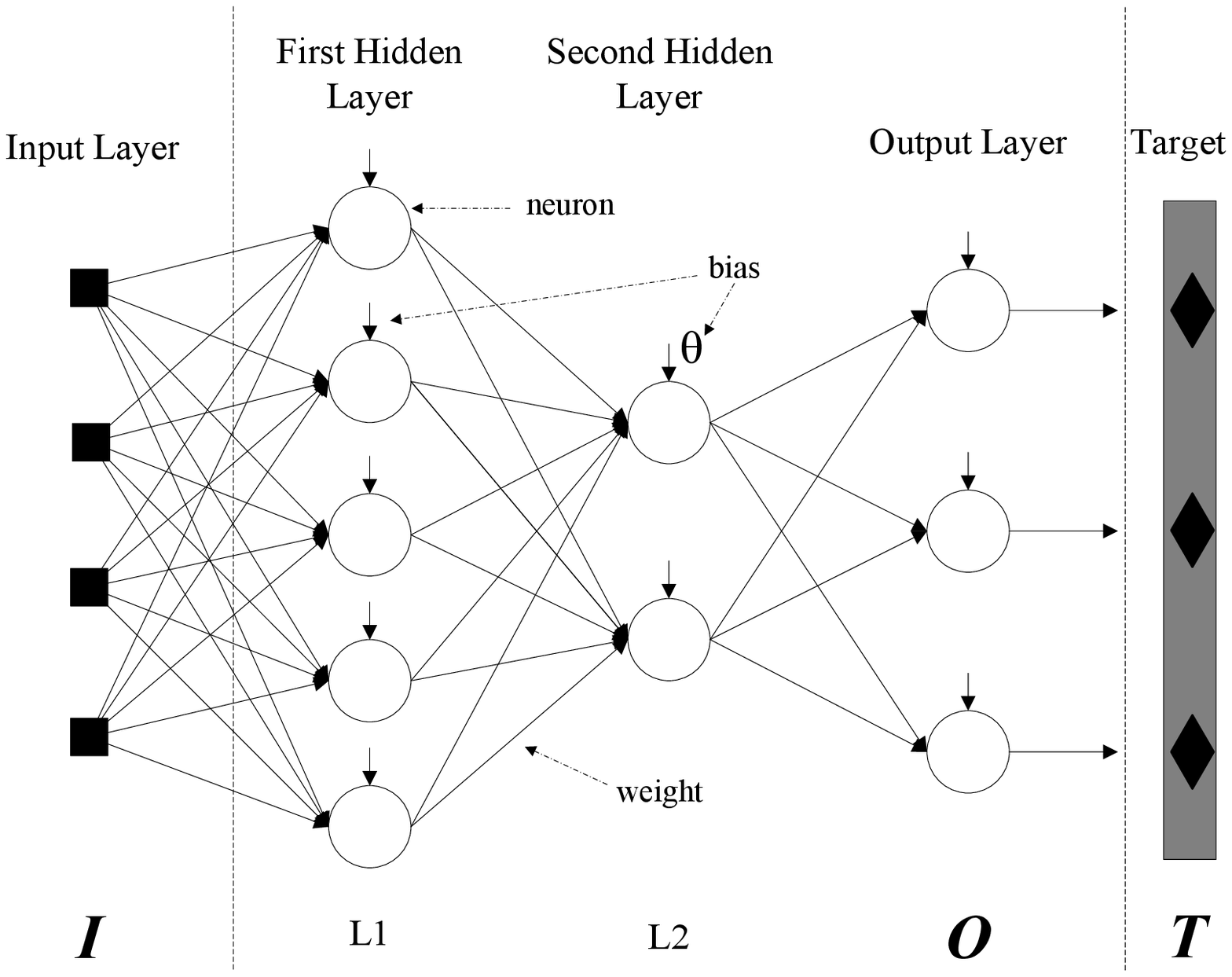} {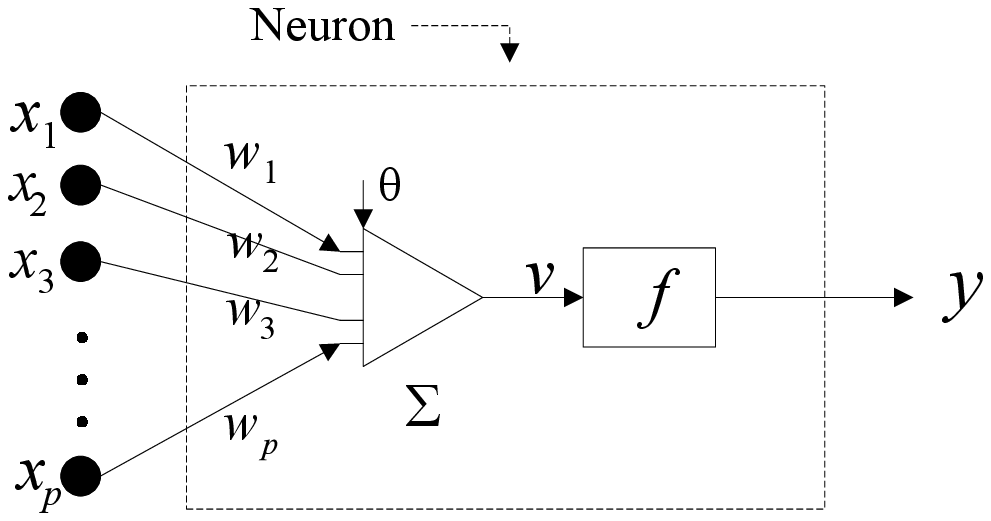}
   \caption{Left: A multilayer feed-forward back propagation network
   with 2 hidden layers. Right: A detailed model of one neuron in BPNN.}
   \label{fA1}
\end{figure}

For clarity and continuity of our work, a BPNN with two hidden
layers is demonstrated below (Fig. \ref{fA1}).
This network contains input layer ($I$ in the figure), hidden layers
($L1$ and $L2$) and output layer ($O$) generally. Input layer reads
training or test patterns (input patterns), which are offered to
be processed by hidden layers and output layer yields the relevant
output results. In our paper, input patterns are corresponding to
5-band magnitudes of cluster and field stars. The desired output
patterns (target patterns) placed at $T$ in Fig.
\ref{fA1} are 1 for cluster stars or 0 for field ones.
The output of BPNN gives the probability of cluster member for each
star.

The network is executed in two phases: training phase and test
phase. In training phase, input and target patterns are submitted to
the network. And then two processes, feeding forward and error back
propagation, are performed. After we endow this network with an initial
state, an input sample (pattern) travels from the input layer. Via being
treated by intermediate layers, the information stored by weights
is processed and a corresponding result comes forth at the output layer.
There, comparing the network output result with the relevant target
pattern, an error performance is calculated. By this error item,
error back propagation is carried out from output to input layer to
modify the connected weights and biases (in Fig.
\ref{fA1}), and store learned knowledge at the same
time. Then, the remain patterns act in the same way and the
iteration goes on until the satisfaction of preplanned error limit.
There are two modes of updating weights and biases: one is
incremental mode in which the weights and biases update when the
errors of patterns back-propagate one by one as presented above and
the other is batch mode in which all patterns travel through the
network and the total error is counted, then the weights and biases
are renewed once in an iteration. We call an iteration of processing
all patterns as one 'epoch'. In test phase, patterns, which have not
been seen by the network, are given to check the efficiency and
accuracy of the network configured in training phase.

The detailed description of the network configuration and concise
mathematics of training it are presented below. In the left panel of
Fig.\ref{fA1}, the input layer and target segment are
divided by dashed lines, which are linked to exoteric environment.
The nodes in the hidden and output layers are called neurons. The
right panel of Fig.\ref{fA1} gives the delicate
structure of one neuron. There are $p$ neurons from the previous
layer as inputs connecting to the neuron enclosed by a dashed
rectangle, where each connection has a weight $w$. In the neuron, an
adder $\sum$ performs to sum all the input values and bias $\theta$
and transmits the result $v$ to a transfer function $f$, which is to
produce an output $y$. Expressions can be presented as
%\begin{mathletters}
\begin{eqnarray}
v &=& \sum_{i=1}^pw_{i}x_i - \theta = \textbf{\textit{w}}'\textbf{\textit{x}} - \theta, \\
y &=& f(v),
\end{eqnarray}
%\end{mathletters}
where $x_i$ is the output of the $i{\rm th}$ node from the previous
layer and $w_{i}$ is the corresponding weight, and
$\textbf{\textit{x}}=[x_1,x_2,\ldots,x_p]',
\textbf{\textit{w}}=[w_1,w_2,\ldots,w_p]'$. More vividly to say, the
stimulus $v$ goes beyond the bias ($\theta$), the neuron will be
activated to release an output signal to next neurons. Here,
$\theta$ can be arranged into the weight vector
$\textbf{\textit{w}}$, as long as we take into account another constant input of
the node. That is to say, we introduce $x_0 = -1$ and
$w_0=\theta$ and let the network adjust $\theta$ just like weights,
so that $v$ has the form of
$v=\textit{\textbf{w}}'\textbf{\textit{x}}$, where
$\textit{\textbf{x}}=[x_0,x_1,\ldots,x_p]',
\textit{\textbf{w}}=[w_0,w_1,\ldots,w_p]'$. Besides, the transfer
function $f$ has various forms, such as:
%\begin{mathletters}
\begin{eqnarray}
\textrm{Linear:}  &f& = v,    \\
\textrm{Log-sigmoid:}  &f& = \frac{1}{1+e^{-v}} \label{eqt-1}, \\
\textrm{Tan-sigmoid:}  &f& = \frac{2}{1+e^{-2v}} - 1,
\end{eqnarray}
%\end{mathletters}
where $-\infty< v <+\infty$. Among these functions, the log-sigmoid
transfer function, which yields results in the range from 0 to 1, is
commonly used in back-propagation networks partly because of its
unlimited differentiability. We will adopt this kind of transfer
functions (Equation \ref{eqt-1}) in our present study.

Now, turn back to training the network. We will select the batch
training mode in this paper. In this mode, weights and bias will be
updated only after the entire inputs and targets are submitted, and
as a result, the gradients (quantitative changes of weights and
bias) are averaged together to produce more accurate estimates. In
this case, a performance function known as mean squared error is used
to evaluate the outputs of the network. The MSE is expressed as a formula
\begin{equation}
E(\textit{\textbf{w}}) =
\frac{1}{N}\sum_{i=1}^{h}(T_i-O_i)^2=\frac{1}{N}(\textbf{T}-\textbf{O})'(\textbf{T}-\textbf{O})=\frac{1}{N}\|\textbf{T}-\textbf{O}\|^2,
\end{equation}
where $N$ is the numbers of input-target pairs, $h$ is the dimension
of output vector $\textit{\textbf{O}}$, $\textit{\textbf{T}}$ is the
target pattern vector, $O_{i}$ and $T_{i}$ are the components of
$\textit{\textbf{O}}$ and $\textit{\textbf{T}}$, and
$\textit{\textbf{w}}$ contains all the weights and bias in the BPNN.
Additionally, target patterns are 1 or 0. Only one output neuron is needed, so $h = 1$ in
this paper.

Given the performance function, a training algorithm should be
provided to teach the BPNN to learn how to classify.
During training, our aim is to make the outputs approach the target
patterns as much as possible, resulting in decreasing the value of
MSE. That is, in order to adjust weights for better learning, we
need to decrease the value of MSE or to minimize this performance
function epoch by epoch. Consequently, an universal training scheme
to update $\textit{\textbf{w}}$ comes up as
\begin{equation}
\triangle\textit{\textbf{w}}^{(k)} =
\textit{\textbf{w}}^{(k+1)}-\textit{\textbf{w}}^{(k)} =
\eta^{(k)}\textit{\textbf{d}}^{(k)}
\end{equation}
or
\begin{equation}
\textit{\textbf{w}}^{(k+1)} = \textit{\textbf{w}}^{(k)} +
\eta^{(k)}\textit{\textbf{d}}^{(k)},
\end{equation}
where $k$ denotes the ${k_{th}}$ epoch of training, the positive
$\eta$ is the learning rate which decides the step length of changes
of $\textit{\textbf{w}}$, and $\textbf{\textit{d}}$ is the search
direction where $\textbf{\textit{w}}$ moves. All the training
algorithms of back propagation network are variations of the above
form. For example, the most basic algorithm, Steepest Descent BP
(SDBP), is based on the negative gradient of
$E(\textit{\textbf{w}})$ as $\textbf{\textit{d}}$. Thus, the
learning rule becomes
\begin{equation}
\triangle\textit{\textbf{w}}^{(k)}=-\eta^{(k)}\nabla{E(\textit{\textbf{w}})}\mid_{\textit{\textbf{w}}=\textit{\textbf{w}}^{(k)}},
\end{equation}
where $\nabla{E(\textit{\textbf{w}})}$ is the differential of $E$.
In this training algorithm, the detailed modifying formula of
$\textit{\textbf{w}}$ in each layer is obtained by chain rule
\citep[$\S$ 11.9]{hn96,hy98}, and relative learning rate can be
optimized \citep[$\S$ 9.6 \& 12.12]{hn96}. There are other
training algorithms to train BPNN: Backpropagation with Momentum
(MOBP) \citep[$\S$ 12.9]{hn96}, Conjugate Gradient Backpropagation
(CGBP) \citep[$\S$ 9.15 \& 12.15]{hn96}, Newton Method using the
Hessian matrix (second derivatives) of the performance as direction
\citep[$\S$ 9.10]{hn96}, Levenberg-Marquardt Backprogation (LMBP)
\citep[$\S$ 12.19]{hn96} and so on. Here, we chose LMBP as our
training algorithm because of its speediest convergence. Detailed algorithm about
LMBP can also be referred to \citet{ba04}.

%% Authors can give a citation as 'Michel et al. 1992'.
%% You may also use \cite, \citep and \citet for citation, and use Table~1 or Figure~1
%% and so forth. Using \ref and \label for cross-references of Tables/Figures
%% is a good way in adjusting/adding/removing text, tables or figures.


\begin{thebibliography}{99}
%% you can type \apj for ApJ, \aap for A&A, \apss for Ap&SS, etc. Please consult
%% the macro chjaa.cls. You can also find them in aasguide.tex (AASTeX for ApJ, AJ, PASP)
%% Please follow the format of ChJAA's reference list
\bibitem[Abazajian et al.(2004)]{ab04} Abazajian, K., et al., 2004, \aj, 128, 502
\bibitem[Adelman-McCarthy et al.(2008)]{ad08} Adelman-McCarthy, J.~K., et al., 2008, \apjs, 175, 297
\bibitem[Bailer-Jones(2000)]{ba00} Bailer-Jones, C.~A.~L., 2000, \aap, 357, 197
\bibitem[Ball et al.(2004)]{ba04} Ball, N.~M., Loveday, J.,Fukugita, M., Nakamura, O., Okamura, S., Brinkmann, J., \& Brunner, R.~J., 2004, \mnras, 348, 1038
\bibitem[Belokurov et al.(2006)]{be06} Belokurov, V., Evans, N. W., Irwin, M. J., Hewett, P. C., \& Wilkinson M. I., 2006, \apj, 637, L29
\bibitem[Bowman et al.(1997)]{bo97} Bowman, A. W., \& Azzalini, A., 1997, Applied Smoothing Techniques for Data Analysis, New York: Oxford University Press
\bibitem[Carney(1984)]{ca84} Carney, B.~W., 1984, \pasp, 96, 841
\bibitem[Folkes et al.(1996)]{fo96} Folkes, S.~R., Lahav, O., \& Maddox, S.~J., 1996, \mnras, 283, 651
\bibitem[Dehnen et al.(2004)]{de04} Dehnen, W., Odenkirchen, M., Grebel, E.~K., \& Rix, H.-W., 2004, \aj, 127, 2753
\bibitem[Grillmair et al.(1995)]{gr95} Grillmair, C. J., Freeman, K. C., Irwin, M., \& Quinn, P. J., 1995, \aj, 109, 2553
\bibitem[Grillmair \& Johnson(2006)]{gr06a} Grillmair, C. J., \& Johnson, R., 2006, \apj, 639, L17
\bibitem[Grillmair \& Dionatos(2006)]{gr06b} Grillmair, C. J., \& Dionatos O., 2006, \apj, 641, L37
\bibitem[Hangan, Demuth \& Beale(1996)]{hn96} Hangan, M.~T., Demuth, H.~B., \& Beale, M.~H., 1996, Neural Network Design, Boston, MA: PWS Publishing
\bibitem[Harris(1996)]{ha96} Harris, W. E., 1996, \aj, 112, 1487
\bibitem[Haykin(1998)]{hy98} Haykin, S., 1998, Neural Networks: A Comprehensive Foundation, 2nd Edition, Harlow, En:Prentice Hall
\bibitem[King(1962)]{ki62} King, I., 1962, \aj, 67, 471
\bibitem[Koch et al.(2004)]{ko04} Koch, A., Grebel, E.~K., Odenkirchen, M., Mart{\'{\i}}nez-Delgado, D., \& Caldwell, J.~A.~R., 2004, \aj, 128, 2274
\bibitem[Lauchner et al.(2006)]{la06} Lauchner, A., Powell JR, W. L., \& Wilhelm, R., 2006, \apj, 651, L33
\bibitem[Leon et al.(2000)]{le00} Leon, S., Meylan, G., \& Combes, F., 2000, \aap, 359, 907
\bibitem[Lupton et al.(2001)]{lu01} Lupton, R., Gunn, J.~E., Ivezi{\'c}, Z., Knapp, G.~R., \& Kent, S., 2001, Astronomical Data Analysis Software and Systems X, 238, 269
\bibitem[Naim et al.(1995)]{na95} Naim, A., et al., 1995, \mnras, 274, 1107
\bibitem[Odewahn et al.(1992)]{od92} Odewahn, S.~C., Stockwell, E.~B., Pennington, R.~L., Humphreys, R.~M., \& Zumach, W.~A., 1992, \aj, 103, 318
\bibitem[Odenkirchen et al.(2001)]{od01} Odenkirche, M., et al., 2001, \apj, 548, L165
\bibitem[Odenkirchen et al.(2003)]{od03} Odenkirche, M., et al., 2003, \aj, 126, 2385
\bibitem[Rockosi et al.(2002)]{ro02} Rockosi, C.~M., et al., 2002, \aj, 124, 349
\bibitem[Schlegel et al.(1998)]{sc98} Schlegel, D.~J., Finkbeiner, D.~P., \& Davis, M., 1998, \apj, 500, 525
\bibitem[Sergios T. et al.(2006)]{se06} Sergios, T., \& Konstantinos, K., 2006, Pattern Recognition, 3rd ed., Athens: Academic Press
\bibitem[Smith et al.(1986)]{sm86} Smith, G.~H., McClure, R.~D., Stetson, P.~B., Hesser, J.~E., \& Bell, R.~A., 1986, \aj, 91, 842
\bibitem[Spitzer(1987)]{sp87} Spitzer, L., 1987, Dynamical Evolution of  Globular Clusters, Princeton, NJ: Princeton University Press
\bibitem[Stoughton et al.(2002)]{st02} Stoughton, C., et al., 2002, \aj, 123, 485
\bibitem[von Hippel et al.(1994)]{hi94} von Hippel, T., Storrie-Lombardi, L.~J., Storrie-Lombardi, M.~C., \& Irwin, M.~J., 1994, \mnras, 269, 97
\bibitem[Wu et al.(2003)]{wu03} Wu, Z.-Y., Shu, C.-G., \& Chen, W.-P., 2003, Chin.Phys.Lett., 20, 1648
\bibitem[Wu et al.(2004)]{wu04} Wu, Z.-Y., Zhou, X., \& Ma, J., 2004, Chin.Phys.Lett., 21, 418

\end{thebibliography}
\end{document}